\def\kms    {\ifmmode{{\rm \ts km\ts s}^{-1}}\else{\ts km\ts s$^{-1}$}\fi}
\def\msol     {\ifmmode{{\rm M}_{\odot}}\else{M$_{\odot}$}\fi}
\def\etal   {{\rm et\ts al.}}
\def\aco  {\ifmmode{^{12}{\rm CO}(J\!=\!1\to0)}\else{$^{12}{\rm CO}(J\!=\!1\to0)$}\fi}
\def\aba {\ifmmode{^{12}{\rm CO}}\else{$^{12}{\rm CO}$}\fi}
\def\abb {\ifmmode{^{13}{\rm CO}}\else{$^{13}{\rm CO}$}\fi}
\def\abc {\ifmmode{{\rm C}^{18}{\rm O}}\else{${\rm C}^{18}{\rm O}$}\fi} 
\def\bco  {\ifmmode{^{12}{\rm CO}(J\!=\!2\to1)}\else{$^{12}{\rm CO}(J\!=\!2\to1)$}\fi}
\def\m    {\ifmmode{\mu {\rm m}}\else{$\mu$m}\fi}
\def\cco  {\ifmmode{^{13}{\rm CO}(J\!=\!1\to0)}\else{$^{13}{\rm CO}(J\!=\!1\to0)$}\fi}
\def\dco  {\ifmmode{^{13}{\rm CO}(J\!=\!2\to1)}\else{$^{13}{\rm CO}(J\!=\!2\to1)$}\fi}
\def\eco  {\ifmmode{{\rm C}^{18}{\rm O}(J\!=\!1\to0)}\else{${\rm C}^{18}{\rm O}(J\!=\!1\to0)$}\fi}
\def\nh   {\ifmmode{N(\hi)}\else{$N$(\hi)}\fi}
\def\hun    {\ifmmode{I_{100}}\else{$I_{100}$}\fi}
\def\sex    {\ifmmode{I_{60}}\else{$I_{60}$}\fi}
\def\hh     {\ifmmode{{\rm H}_2}\else{H$_2$}\fi}
\def\nhh     {\ifmmode{N({\rm H}_2)}\else{$N$(H$_2$)}\fi}
\def\dhh     {\ifmmode{n({\rm H}_2)}\else{$n$(H$_2$)}\fi}
\def\zwco   {\ifmmode{^{12}{\rm CO}}\else{$^{12}{\rm CO}$}\fi}
\def\nzwco   {\ifmmode{N(^{12}{\rm CO})}\else{$N(^{12}{\rm CO})$}\fi}
\def\wzwco   {\ifmmode{W(^{12}{\rm CO})}\else{$W(^{12}{\rm CO})$}\fi}
\def\drco   {\ifmmode{^{13}{\rm CO}}\else{$^{13}{\rm CO}$}\fi}
\def\ndrco   {\ifmmode{N(^{13}{\rm CO})}\else{$N(^{13}{\rm CO})$}\fi}
\def\wdrco   {\ifmmode{W(^{13}{\rm CO})}\else{$W(^{13}{\rm CO})$}\fi}
\def\tex    {\ifmmode{T_{ex}({\rm CO})}\else{$T_{ex}({\rm CO})$}\fi}
\def\tkin    {\ifmmode{T_{kin}}\else{$T_{kin}$}\fi}
\def\xco     {\ifmmode{X_{\rm CO}}\else{$X_{\rm CO}$}\fi} 
\def\ha     {\ifmmode{{\rm H}\alpha}\else{${\rm H}\alpha$}\fi}

\newcommand{\hi}{H\,{\small{\sc I}}}

\documentclass[here,epsfig]{aa}
\usepackage{graphics}
\begin{document}

   \thesaurus{03  
              (09.05.1;  
               09.13.2;  
               09.19.1;  
               11.09.1;  
               11.09.4;  
               11.19.3)} 

\title{The effect of violent star formation on the state of the molecular gas in M\,82} 


   \author{A. Wei\ss\inst{1} N. Neininger\inst{1} S. H\"uttemeister\inst{1}
          \and U. Klein\inst{1}}

   \offprints{A. Wei\ss}

   \institute{Radioastronomisches Institut der Universit\"at Bonn (RAIUB),
              Auf dem H\"ugel 71, D-53121 Bonn, Germany\\
              email: aweiss@astro.uni-bonn.de}

   \date{Received July 27, 2000; accepted October 11, 2000}
   \authorrunning{A. Wei\ss\, \etal} 

   \titlerunning{The state of the molecular gas in M\,82}
   \maketitle

   \begin{abstract}

We present the results of a high angular resolution, multi--transition 
analysis of the molecular gas in M\,82. The analysis is based on the two 
lowest transitions of $^{12}{\rm CO}$ and the ground transition of the rare 
isotopes $^{13}{\rm CO}$ and ${\rm C}^{18}{\rm O}$ measured with the PdBI, 
the BIMA array and the IRAM 30m telescope.\\
In order to address the question of how the intrinsic molecular cloud 
properties are influenced by massive star formation we have carried out 
radiative transfer calculations based on the observed CO line ratios. 
The calculations suggest that the kinetic temperature of the molecular 
gas is high in regions with strong star formation and drops towards the 
outer molecular lobes with less ongoing star formation. The location of 
the highest kinetic temperature is coincident with that of 
the mid infrared (MIR) peaks which trace emission from hot dust. The hot 
gas is associated with low H$_2$ densities while the cold gas in the outer 
molecular lobes has high H$_2$ densities. We find that CO intensities do 
not trace \hh\, column densities well. Most of the molecular gas is distributed 
in a double--lobed distribution which surrounds the starburst. A detailed 
analysis of the conversion factor from CO intensity to \hh\, column density 
shows that \xco\, depends on the excitation conditions. We find $\xco\, \sim 
{\rm T}_{\rm kin}^{-1}\,\,{\rm n(\hh)}^{1/2}$, as expected for virialized clouds.

   \end{abstract}

%

\section{Introduction}

M\,82 is regarded as the archetypical starburst galaxy (Rieke \etal\ 
\cite{rieke}). Its distance of only 3.9\,Mpc (Sakai \& Madore 
\cite{sakai}) makes M\,82 an excellent laboratory for studying the relevant 
physical processes connected with starburst activity in detail. The central 
few hundred parsecs of this galaxy are heavily obscured by dust and gas which 
hides the central starburst region against direct observations at optical 
wavelengths. Evidence for the strong star--forming activity in the central 
region comes from radio (e.g. Kronberg \etal\ \cite{kronberg}; Wills \etal\ 
\cite{wills99}) and infrared observations (e.g. Telesco \& Gezari \cite{telesco92})
and also from the prominent bipolar outflow visible in \ha\,(e.g. Bland \& 
Tully \cite{bland}; McKeith \etal\ \cite{mckeith}; Shopbell \& Bland--Hawthorn 
\cite{shopbell}) and X--rays (e.g. Bregman \etal\ \cite{bregman}). The massive 
star formation (SF) is believed to be fueled by the large amount of molecular 
gas which is present in the centre of M\,82.\\
On the other hand, SF affects the distribution, kinematics and physical 
conditions of the surrounding interstellar medium (ISM). Early studies of 
the distribution of the molecular gas in M\,82 unveiled a double--lobed 
circumnuclear distribution of CO which was interpreted as a molecular torus 
with a depletion of molecular gas in the central region (Nakai \etal\ 
\cite{nakai}). More recent high--resolution studies by Shen \& Lo (\cite{shen}) 
using BIMA and Neininger \etal\ (\cite{nico}) using the IRAM interferometer at 
Plateau de Bure (PdBI) showed a third molecular peak 65\,pc west of the 2.2\m\, 
nucleus (Dietz \etal\ \cite{dietz}). 
Using these high--resolution CO maps Wei\ss\, \etal\ (\cite{aweiss}) identified 
an expanding superbubble in the molecular gas of M\,82 which linked the triple 
peak CO distribution and its disturbed velocity field to the prominent outflow
visible in \ha\, and X--rays.\\
Multi--transition analyses of molecular emission lines (CO, CS, HCN) showed 
that the starburst also affects the physical conditions of the molecular gas 
(Wild \etal\ \cite{wild}; Henkel \& Bally \cite{henkel85}; Brouillet \& 
Schilke \cite{brouillet}). A large fraction of the molecular gas is concentrated 
in warm (\tkin\,=\,50\,K) and dense (\dhh\,=\,10$^4$\,cm$^{-3}$) clouds (Wild 
\etal\ \cite{wild}; G\"usten \etal\ \cite{guesten}). In a recent study Mao 
\etal\ (\cite{mao}) analyzed CO mm and sub--mm emission lines in M\,82 to 
investigate the physical properties of the molecular clouds. They conclude that 
the bulk of CO emission arises from photon--dominated regions (PDRs) while 
tracers of high--density gas like CS and HCN are less affected by the strong 
UV radiation from massive stars.\\
Even though these studies already provided a good global picture of the physical 
conditions of the molecular gas in M\,82, no detailed high--resolution study 
exists so far that allows to resolve variations of the excitation conditions 
of the molecular gas over the central part of M\,82. In this paper we present 
the results of a high angular resolution, multi--transition CO analysis and 
compare the intrinsic gas properties with observations of high--level star 
formation. In Sect.\,2 we briefly summarize our observations and the data 
reduction. In Sect.\,3 we describe the main results including a description 
of the CO morphology and kinematics, the CO line ratios, results from the 
LVG calculation and on \xco. In Sect.\,4 we compare our results to previous 
studies. Sect.\,5 summarizes our conclusions.


\section{Observations}

\subsection{PdBI Observations}

We have used the PdBI to observe the
\bco\, ($\nu_{obs}\,=\,230.361$\,GHz) and \eco\, ($\nu_{obs}\,=\,109.698$\,GHz) 
emission lines in the central region of M\,82. The observations were carried out 
in April 1997. Due to the dual frequency setup of the PdBI we were able to 
observe both emission lines simultaneously. The observations were carried out in 
mosaic mode with seven pointings covering the central kpc of M\,82. The central 
pointing was centred on the 2.2\m\, nucleus at $\alpha = 09^h55^m51^s.94, 
\delta = 69^{\circ}40'47.14''$ (J2000.0) (Dietz \etal\ \cite{dietz}). 
The other pointings were shifted with 
respect to the central position by $(\alpha,\delta)$\,=\,$(-30'',-8'')$,\,
$(-20'',-6'')$,\,$(-10'',-4'')$,\,$(10'',4'')$,\,$(20'',8'')$,\,$(30'',10'')$ 
which ensured sufficient overlap of the observed fields at 230\,GHz. The primary 
beam of the PdBI is $22''$ and $45''$ at 230\,GHz and 109\,GHz respectively. 
The observations were carried out in the DC2 antenna configuration with 
baselines ranging from 24\,m to 176\,m leading to a synthesized beam of 
$3.8''\times 3.5''$ at 109\,GHz and $1.5''\times 1.4''$ at 230\,GHz. \\
The \eco\, data were recorded using two correlator units leading to a total 
bandwidth of 780\,\kms\, with 6.83\,\kms\, resolution. For the \aco\, transition 
we used four correlator units which resulted in a total bandwidth of 390\,\kms\, 
and a velocity resolution of 3.25\,\kms. The \bco\, emission line was observed 
in the lower sideband of the 230\,GHz, the \eco\, emission line in the upper 
sideband of the 109\,GHz receiver. The flux and complex bandpass calibration 
was determined by observing 3C273 and MWC349. The nearby calibrator 0836+710 
was used as a secondary amplitude and phase calibrator. The seven fields were
combined in a mosaic and subsequently cleaned using the MAPPING procedure of the 
GILDAS software package. This yields a roughly constant sensitivity along the 
major axis of M\,82 with an rms noise of 6\,mJy/beam at 109\,GHz and 
30\,mJy/beam at 230\,GHz. For both data sets the channels with 
$v_{lsr} > 385\,\kms$ and $v_{lsr} < 30\,\kms$ were used to generate a continuum 
map at 109\,GHz and 230\,GHz. The continuum emission was subtracted from both
emission line data cubes.    

\subsection{IRAM 30m Telescope Observations}

In addition to the high--resolution CO data we observed the \aco, \bco\, and 
\cco\, emission lines with the IRAM 30m telescope in on--the--fly mode. The 
observations covered an area of $3'\times3'$ centred on the 2.2\m\, nucleus. 
The \aco\, observations were carried out in Nov 1997. The \bco\, and \cco\, 
data  were observed in Nov. 1997, Dec. 1998 and June 1999. For all observations 
we used the same observing procedure: The scanning velocity was $2''$/sec and 
the readout sampling 1\,sec leading to a spatial separation of $2''$ between 
individual spectra in scanning direction. The spatial separation between 
individual scans was $4''$. Thus each on--the--fly map was sampled on 
a $2''\times 4''$ grid. For the \aco\, transition we performed two coverages, 
for the other two transitions we performed four coverages with perpendicular 
scanning directions. The combined data therefore were sampled on a $2''\times 2''$ 
grid. After first--order baseline subtraction the spectra were summed on a 
$3''\times 3''$ grid using the beam ($13''$ at 230\,GHz, $22''$ at 115\,GHz) 
and the rms noise level for weighting. This observing and reduction procedure 
guarantees a smooth data sampling and avoids artifacts in the combination with 
the interferometric data. The total integration time per beam was 65\,sec for 
the \aco, 45\,sec for the \bco\, and 130\,sec for \cco\, transition resulting 
in an rms noise of 40\,mK, 65\,mK and 15\,mK. As backends we used the 
autocorrelators which lead to a total bandwidth and velocity resolution of 
650\,\kms/2.6\,\kms, 650\,\kms/1.3\,\kms and 695\,\kms/2.7\,\kms for the \aco, 
\bco\, and \cco\, transitions respectively. For the conversion from $T_A^*$ to 
$T_{mb}$ we used $F_{eff}/B_{eff}=1.35$ at 115\,GHz (Gu\`elin \etal\ 
\cite{guelin}) and $F_{eff}/B_{eff}=2.05$ at 230\,GHz (Greve \etal\ 
\cite{greve}). 

\subsection{Short--Spacing Correction}

To ensure that the interferometric line intensities do not suffer from 
missing flux due to extended emission we combined the interferometer and 
the single--dish data cubes. For the combination we used a method that 
works on the final reduced (CLEANed and corrected for primary beam attenuation) 
interferometer cubes. The only free parameter in this method is the choice 
of which part of the {\it uv}-plane in the interferometer cube is replaced 
by the single--dish values.  A detailed description of the method is given 
in appendix A. The parameters for the 30m beam sizes, the corresponding 
effective diameter of the 30m telescope, the shortest baseline, the replaced 
part of the {\it uv}--plane and the missing flux of the interferometer maps 
are given in Table \ref{ssc}. All reduction steps were done using the MIRIAD 
software package. We applied the short--spacing correction to the \aco\, 
cube obtained by Shen \& Lo (\cite{shen}), the \cco\, cube from Neininger \etal\ 
(\cite{nico}) and to the \bco\, cube.  

\begin{table}
\caption{Summary of the relevant parameters of the short--spacing correction.}
\begin{tabular}{l c c c}
& $^{12}$CO(1--0) &$^{12}$CO(2--1) &$^{13}$CO(1--0) \\ \hline
FWHM & 22$''$ &  11$''$&  22$''$\\
$D_{eff}$ & 28.1\,m & 24.4\,m &  28.1\,m \\
$S_{min}$ & unknown &  24\,m & 24\,m \\
SD[$\frac{D_{eff}}{k\lambda}$] & $<$9.4 &  $<$18.5& $<$8.0 \\
miss. flux & 20 \% &  60 \% &  35 \%\\
\end{tabular}
\vspace*{0.2cm}\\
 From top to bottom the parameters are the FWHM of the IRAM 30\,m telescope beam, 
the corresponding effective area of the 30\,m telescope, the shortest 
baseline in the interferometer data, the part of the {\it uv}--plane that 
has been replaced by the single--dish data and the average missing flux 
in the interferometer maps. Since the shortest baseline of the \aco\, 
observations from Shen \& Lo is not given in their paper we replaced the part 
of the {\it uv}--plane which corresponds to the effective diameter of the 
30\,m telescope.\label{ssc}
\end{table}

\section{Results}

\subsection{CO morphology and kinematics}

\begin{center}
\begin{figure*}
\resizebox{14cm}{!}{\includegraphics{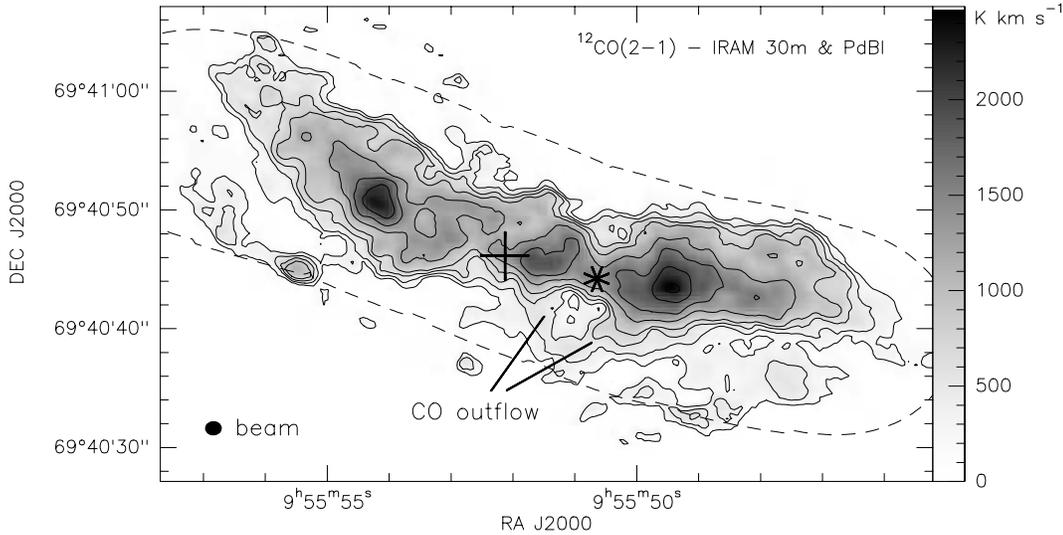}}
\caption{Integrated \bco\, line intensity derived from the short--spacing 
corrected data cube. The contours correspond to 200, 400, 600, 800, 1200, 
1600, 2000 and 2400 K\kms. The cross indicates the centre of M\,82 
(2.2\m\, peak), the star represents the location of the brightest radio 
continuum point source (SNR 41.9+58) in M\,82. The dotted line shows the 
50\% sensitivity level of the primary beam mosaic.}
\label{m0-12co21-ssc}
\end{figure*}
\end{center}

\begin{center}
\begin{figure*}
\resizebox{14cm}{!}{\includegraphics{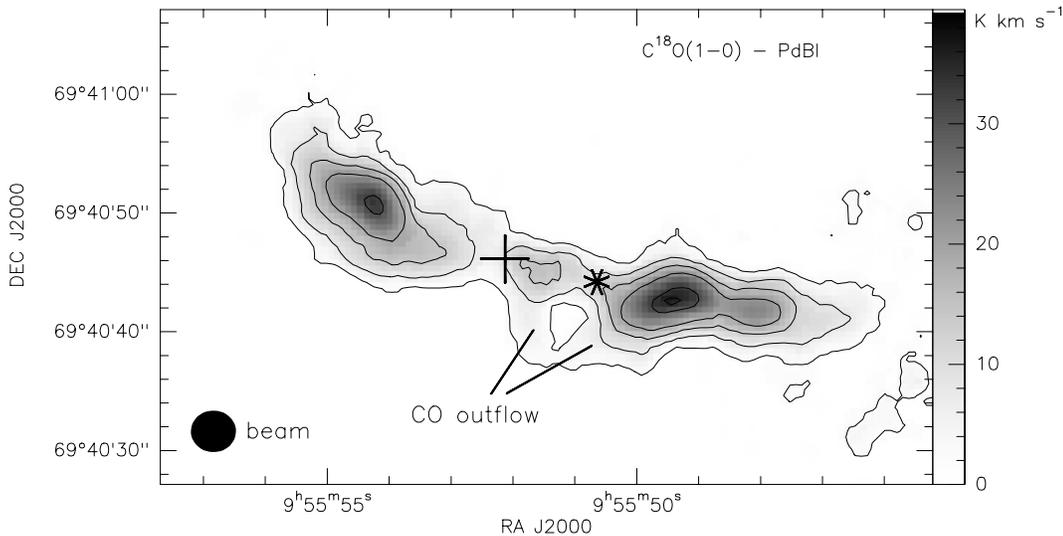}}
\caption{Integrated \eco\, line intensity. The contours correspond
to 2, 8, 14, 20, 26 and 32 K\kms. The cross indicates the centre 
of M\,82 (2.2\m\, peak), the star marks the location of the brightest radio 
continuum point source (SNR 41.9+58) in M\,82.}
\label{m0-c18o}
\end{figure*}
\end{center}

\begin{figure}
\resizebox{9cm}{!}{\includegraphics{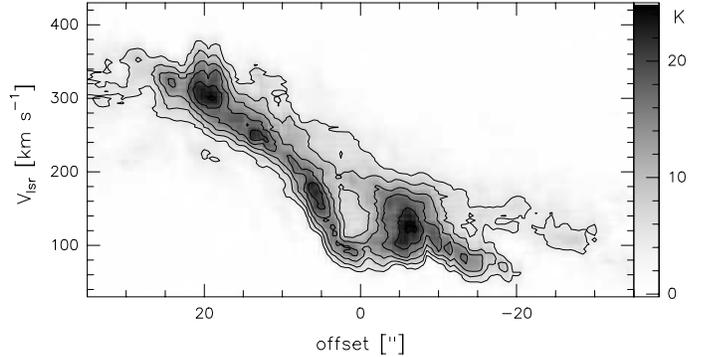}}
\caption{A pv--diagram along the major axis of M\,82 in the 
\bco\, transition. The contours correspond to 4, 8, 12, 14, 18, 
and 22 K. The pv--diagram is centred on the brightest radio 
continuum point source SNR 41.9+58 (see Figs.\ \ref{m0-12co21-ssc} 
\& \ref{m0-c18o}). The velocity crowded regimes at $20''$, $5''$ and $-7''$ 
offset correspond to the western, central and eastern CO peak.}
\label{pv-12co21}
\end{figure}

\begin{figure*}
\resizebox{17cm}{!}{\includegraphics{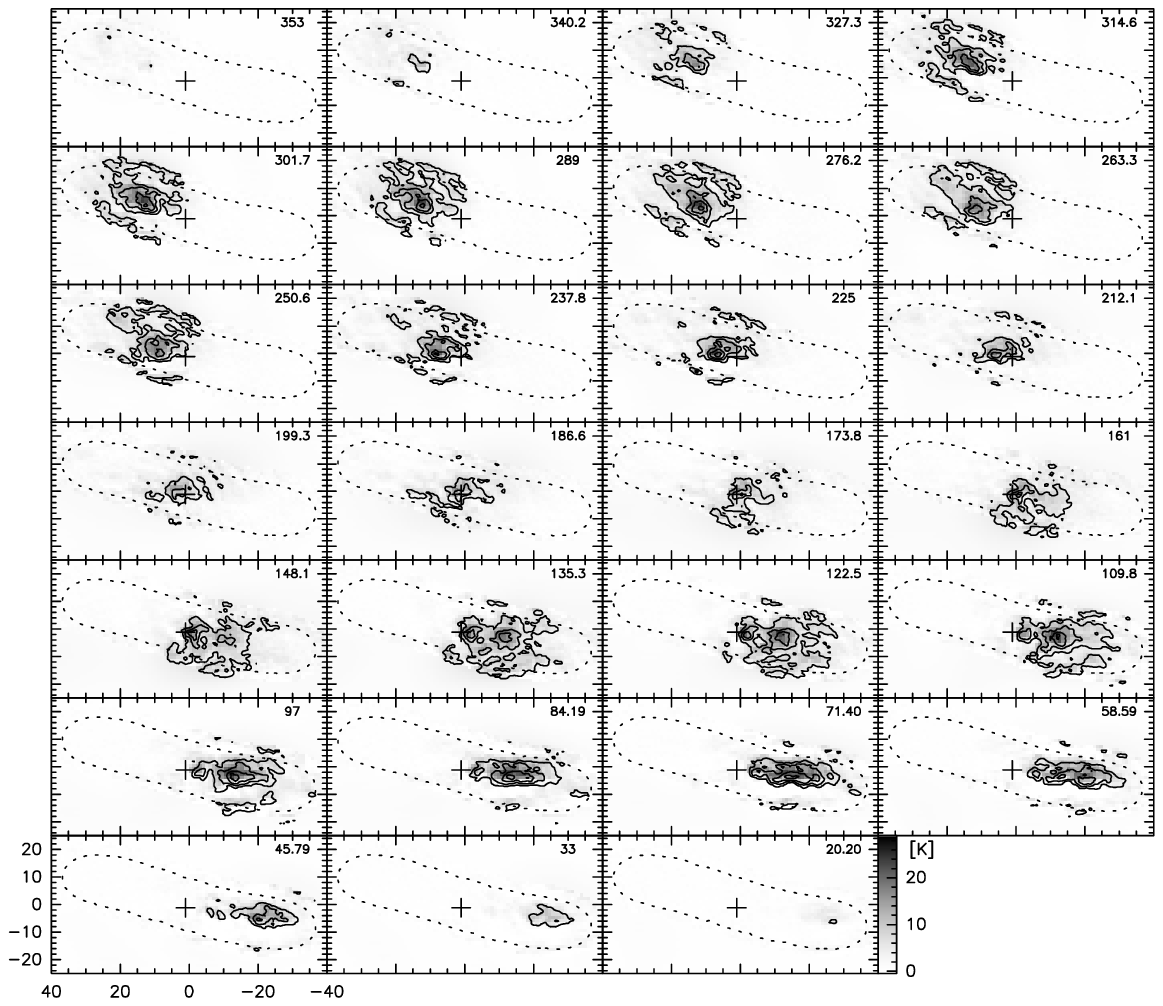}}
\caption{\bco\, velocity channel maps of the short--spacing corrected 
data cube. Offsets in RA, DEC are given relative to the 2.2\m\, nucleus 
indicated by the cross in each channel map. The velocity spacing between 
individual maps is 12.8\,\kms, the central velocity ($V_{lsr}$) of each 
map is given in the top right corner in units of \kms. The intensities 
are corrected for attenuation of the primary beam. The contours correspond 
to 6, 12, 18, 24 and 30 K. The dotted line in each channel map shows the 50\% 
sensitivity level from the primary beam mosaic.}
\label{channels12}
\end{figure*}

\begin{figure*}
\resizebox{17cm}{!}{\includegraphics{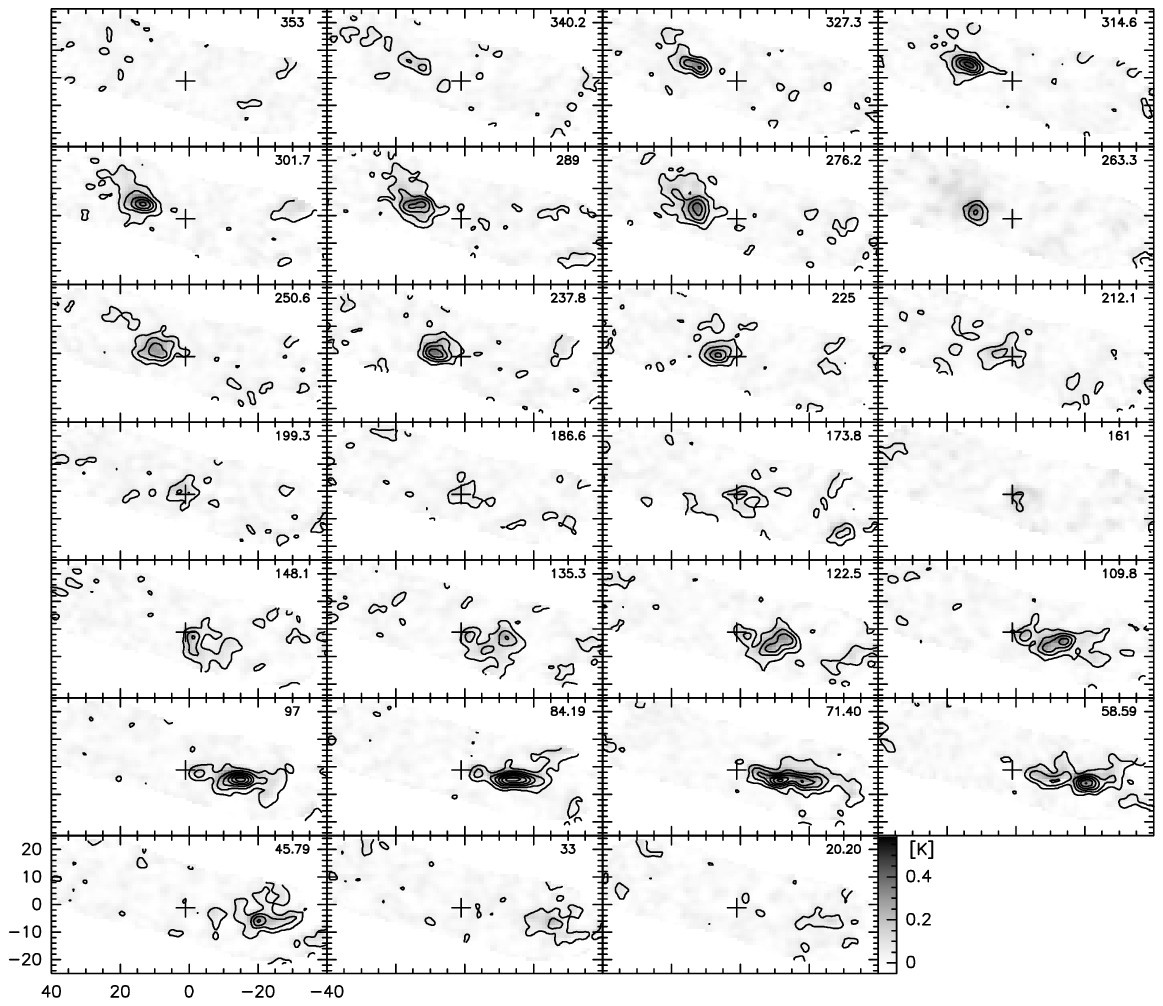}}
\caption{\eco\, velocity channel maps of the PdBI data cube. The figure has
the same layout and specifications as Fig.\ \ref{channels12} 
Contours correspond to 0.05, 0.15, 0.25, 0.35, 0.45 and 0.55 K.}
\label{channels18}
\end{figure*}

Figs. \ref{m0-12co21-ssc} and \ref{m0-c18o} show the integrated \bco\, and \eco\, 
line intensities. The overall morphology of both images is very similar to the 
\aco\, distribution published by Shen \& Lo (\cite{shen}) and the \cco\, 
distribution published by Neininger \etal\ (\cite{nico}). It shows a 
triple peak morphology of which the two outer lobes have been interpreted as the 
edge of a central molecular toroid (Nakai \etal\ \cite{nakai}, Shen \& Lo 
\cite{shen}) and a weaker central peak located 65\,pc west of the M\,82`s 
centre (2.2\m\, peak; Dietz \etal\ \cite{dietz}). The two outer lobes 
have a projected separation of 410\,pc (26$''$). The separation of the central 
and the western molecular lobe is only about 130\,pc (8$''$). More diffuse CO 
emission is detected in the \bco\, intensity distribution east and west of the 
CO peaks and in the south--west of the galaxy. The eastern part of the CO 
distribution is significantly warped to the north. The total extent of the 
emission region is about 1\,kpc from east to west. With respect to M\,82`s 
centre the distribution of the molecular gas is clearly displaced to the west. 
South of the central and western CO peak two CO spurs are detected 
(see Fig\,\ref{m0-12co21-ssc}). They extend about 100\,pc below the main 
molecular disk and join just below the expanding molecular superbubble which 
is located between the central and western CO peak (Neininger \etal\ \cite{nico}; 
Wei\ss\, \etal\ \cite{aweiss}). At the same location hot gas emerges into the 
halo of M\,82 (e.g.  Shopbell \& Bland--Hawthorn \cite{shopbell}; Bregman \etal\ 
\cite{bregman}) supporting the idea that the CO spurs indicate the walls of 
the superbubble. \\ 
Note that the chain of CO emission south of the eastern end of the \bco\, 
distribution is most likely not real but an artifact from the primary beam 
correction.\\ The kinematic of the central 400\,pc is dominated by solid 
body rotation. The rotation amplitude is about 200\,\kms\, ranging from 
115\,\kms\, at the western peak up to 320\,\kms\, at the eastern peak. A 
pv--diagram along the major axis of M\,82 in the \bco\, transition is shown 
in Fig. \ref{pv-12co21}. (For the corresponding diagram in the \eco\, data 
see Wei\ss\, \etal\ \cite{aweiss}). The pv-diagram is centred on the 
brightest supernova remnant SNR 41.9+58. The intense, velocity crowded regions at 
$20''$, $5''$ and $-7''$ offset correspond to the western, central and 
eastern CO peak. Between the central and western CO peaks two velocity 
components at 100\,\kms\, and 190\,\kms\, are detected. These features 
have been interpreted as an expanding superbubble. The velocity of the CO spurs 
is about 140\,\kms (see Figs. \ref{channels12} and \ref{channels18}) which is 
similar to the centroid velocity of the expanding superbubble. Outside the 
central 400\,pc the CO rotation curve flattens. The dynamical centre derived 
from the \bco\, and \eco\, data agrees very well with the value of 
${\rm v}_{sys}=225\pm10\,\kms$ published by Shen \& Lo (\cite{shen}) for 
the \aco, and Neininger \etal\ (\cite{nico}) for the \cco\, transition. 
The channelmaps of the \bco\, and \eco\, line emission are presented in 
Figs. \ref{channels12} and \ref{channels18}. 

\subsection{Location of the starburst with respect to CO}

Most tracers of star formation in M\,82 indicate that the highest 
star--forming activity is not associated with the molecular peaks, 
which presumably indicate the location of the reservoirs for the `fuel' 
for star formation, but rather takes place between the peaks. 
The high--resolution 12.4 \m\, image of the central region of 
M\,82 published by Telesco \& Gezari (\cite{telesco92}) suggests that 
the young stellar clusters, which heat the dust, are located between the 
western molecular lobe and the 2.2 \m\, nucleus (western mid infrared (MIR) 
peaks), at the central CO peak, and between the central CO peak and the eastern 
CO lobe (eastern MIR peak). A similar morphology is visible in the Ne{\small II} 
line emission (Achtermann \& Lacy \cite{achtermann}). The radio continuum point 
sources, which are believed to be supernova remnants (SNR) and compact 
H{\small II} regions, are spread across a much wider region and seem to avoid 
MIR and Ne{\small II} peaks (Kronberg \etal\ \cite{kronberg}). Only the 
strongest SNR in M\,82, SNR41.9+58, appears to be related to features at 
other wavelengths: it is located near the centre of the expanding molecular 
superbubble, between the central and western CO peak, from which hot X--ray 
emitting gas is released into the halo of M\,82 (Wei{\ss} \etal\ \cite{aweiss}). 
 At the same location recent radio continuum studies by Wills \etal\ 
(\cite{wills99}) identified a blow--out in the form of a cone of missing 5--GHz 
continuum emission. In the same study three other chimneys were identified 
within the central 300\,pc of M\,82. All these observations 
indicate that the regions of violent star formation are confined by the 
molecular lobes. Since no indications for high activity have been found at the 
2.2 \m\, nucleus itself, it seems that the starburst is arranged in a toroidal 
topology around the nucleus.

\subsection{CO line ratios}

To calculate the line ratios properly we used the short--spacing corrected 
\aco,\bco\, and \cco\, data cubes. Note that the missing flux in the pure 
interferometric maps can be as high as 60\% (see Table \ref{ssc}). Therefore 
the short--spacing correction is vital to derive proper line ratios. The
short--spacing correction is less crucial for the peak intensities. 
Here the missing flux is 10\%--30\% only. The \aco, \bco\, and \eco\, data 
were smoothed to the resolution of the \cco\, observations ($4.2''$). 
Since no single--dish data were obtained for the \eco\, transition 
we applied the missing flux factors derived from the \cco\, peak intensity 
distribution to the \eco\, observations. This procedure is justified, 
because the frequency of both transitions is similar and the observations 
were carried out in the same configurations with the PdBI. This leads to 
similar {\it uv}--coverages for both observations. Furthermore the 
morphology in the interferometer maps is similar and both transitions are 
optically thin (see Sect. \ref{lvg}). To take the 
remaining uncertainties into account we assumed an error of 50\% for the \eco\, 
line intensities. The line ratios were calculated at 19 positions across the CO 
distribution of M\,82. The spacing between individual positions is about $4''$. 
The analyzed positions are marked by the crosses in Fig. \ref{lvg-para}. 
The circles indicate the FWHM of $4.2''$ used in the study. The positions 
include all molecular peaks, the 2.2\m\, nucleus, the MIR peaks, the CO spurs 
and the diffuse emission in the outer regions of M\,82.  For clarity the 
positions have been labeled 1 to 16 from east to west. Positions 17 to 19 
correspond to positions on the CO outflow (see Fig.\ \ref{lvg-para}). 
The line ratios at the analyzed positions are summarized in Tab~\ref{lineratios}. 
Errors include 10\% uncertainty of the flux calibrators, errors of the 
amplitude calibration (typically about 10\%) and statistical errors. Our 
high--resolution line ratios for \aba\, and \abb\, differ slightly
from values derived from single dish observations by Mao \etal\ (\cite{mao}).
But our data confirms that \bco/\aco\, ratios larger than 1.8 
(e.g. Knapp \etal\ \cite{knapp}, Olofsson \& Rydbeck \cite{olofsson}, Loiseau 
\etal\ \cite{loiseau}) can firmly be rejected. \aco/\cco\, and \aco/\eco\, 
line intensity ratios are about 10--20 and 40--60 respectively.

\begin{table*}
\caption{CO line ratios in M\,82 at 4$''$ resolution. The offsets are given 
relative to $\alpha = 09^h55^m51^s.94, \delta = 69^{\circ}40`47.14''$ (J2000.0). 
For Column\,8 \& 9 we have corrected the C$^{18}$O intensities with the missing 
flux factor determined from the $^{13}$CO data. The corresponding percentage 
is given in Column\,10. Errors include 10\% uncertainty of the flux calibrators, 
errors of the amplitude calibration (typically about 10\%) and statistical 
errors. For the C$^{18}$O intensities we assumed an error of 50\% due to the 
unknown missing flux. \label{lineratios}}

\begin{tabular}{c c c c c c c c c c}
&$\Delta \alpha$ & $\Delta \delta$
&$\frac{{\rm I(CO}(2-1))}{{\rm I(CO}(1-0))}$
&$\frac{{\rm T(CO}(2-1))}{{\rm T(CO}(1-0))}$
&$\frac{{\rm I(CO}(1-0))}{{\rm I(^{13}CO}(1-0))}$
&$\frac{{\rm T(CO}(1-0))}{{\rm T(^{13}CO}(1-0))}$
&$\frac{{\rm I(CO}(1-0))}{{\rm I(C^{18}O}(1-0))}$
&$\frac{{\rm T(CO}(1-0))}{{\rm T(C^{18}O}(1-0))}$
&MF$^{13}{\rm CO}^1$\\
&$['']$ & $['']$ & & & & & & &[\%] \\ \hline
1& 16.5& 7.5&$1.16\pm0.3$&$1.02\pm0.3$&$15.2\pm4.9$&$19.2\pm5.7$&$44\pm24$&
$66\pm35$&23.5\\
2& 14.5& 5.0&$1.17\pm0.3$&$0.96\pm0.3$&$12.7\pm3.9$&$15.4\pm4.9$&$38\pm21$&
$45\pm24$&10.3\\
3& 11.5& 3.0&$1.12\pm0.3$&$0.98\pm0.3$&$11.5\pm3.4$&$12.2\pm3.7$&$40\pm21$&
$44\pm24$&8.0\\
4&  9.5& 1.0&$1.19\pm0.3$&$1.09\pm0.3$&$12.0\pm3.6$&$13.0\pm4.1$&$39\pm21$&
$43\pm23$&7.0\\
5&  6.5& 0.5&$1.34\pm0.4$&$1.03\pm0.3$&$14.2\pm4.4$&$12.8\pm4.2$&$37\pm20$&
$40\pm22$&9.0\\
6&  4.0& 0.5&$1.42\pm0.4$&$1.28\pm0.4$&$15.0\pm4.8$&$12.6\pm4.1$&$33\pm18$&
$37\pm20$&12.4\\
7&  2.0&-0.5&$1.39\pm0.4$&$1.26\pm0.4$&$17.3\pm5.7$&$16.5\pm6.1$&$37\pm20$&
$61\pm33$&10.3\\
8& -1.0&-1.5&$1.06\pm0.3$&$1.01\pm0.3$&$25.9\pm8.5$&$20.7\pm8.0$&$36\pm20$&
$56\pm30$&13.0\\
9& -4.0&-2.0&$0.98\pm0.3$&$0.98\pm0.3$&$21.3\pm6.6$&$22.1\pm8.4$&$53\pm28$&
$56\pm30$&15.6\\
10& -6.5&-3.0&$1.07\pm0.3$&$0.95\pm0.3$&$25.4\pm8.4$&$17.0\pm6.0$&$41\pm22$&
$45\pm24$&12.9\\
11&-10.0&-4.0&$1.12\pm0.3$&$1.08\pm0.3$&$13.2\pm3.9$&$11.7\pm3.8$&$45\pm24$&
$29\pm15$&14.0\\
12&-14.0&-4.0&$1.00\pm0.3$&$0.98\pm0.3$&$15.3\pm4.5$&$11.5\pm3.5$&$48\pm26$&
$36\pm20$&4.3\\
13&-17.5&-4.0&$1.14\pm0.3$&$1.02\pm0.3$&$14.6\pm4.4$&$10.9\pm3.3$&$48\pm26$&
$39\pm21$&6.6\\
14&-20.5&-4.5&$1.05\pm0.3$&$0.95\pm0.3$&$14.3\pm4.3$&$11.1\pm3.6$&$49\pm27$&
$39\pm21$&4.1\\
15&-23.5&-4.5&$1.07\pm0.3$&$1.00\pm0.3$&$16.3\pm5.1$&$14.7\pm4.7$&$60\pm32$&
$48\pm26$&5.6\\
16&-26.5&-4.0&$1.14\pm0.3$&$1.09\pm0.3$&$14.5\pm4.8$&$14.3\pm4.9$&$62\pm34$&
$56\pm30$&6.2\\ \hline
17& -2.5&-5.5&$1.02\pm0.3$&$1.07\pm0.3$&$20.2\pm6.8$&$21.2\pm8.3$&$40\pm22$&
$54\pm29$&19.0\\
18& -3.5&-9.0&$1.20\pm0.3$&$0.91\pm0.3$&$12.0\pm4.1$&$21.5\pm8.6$&$29\pm16$&
$52\pm28$&27.4\\
19& -8.5&-7.0&$1.36\pm0.4$&$1.12\pm0.3$&$11.8\pm3.8$&$11.6\pm3.9$&$32\pm17$&
$33\pm18$&10.8\\
\end{tabular}
\vspace*{0.2cm}

$^1$Missing flux determined from the $^{13}$CO data
\end{table*}

\subsection{Radiative transfer calculations \label{lvg}}

\begin{center}
\begin{figure*}
\resizebox{14.0cm}{!}{\includegraphics{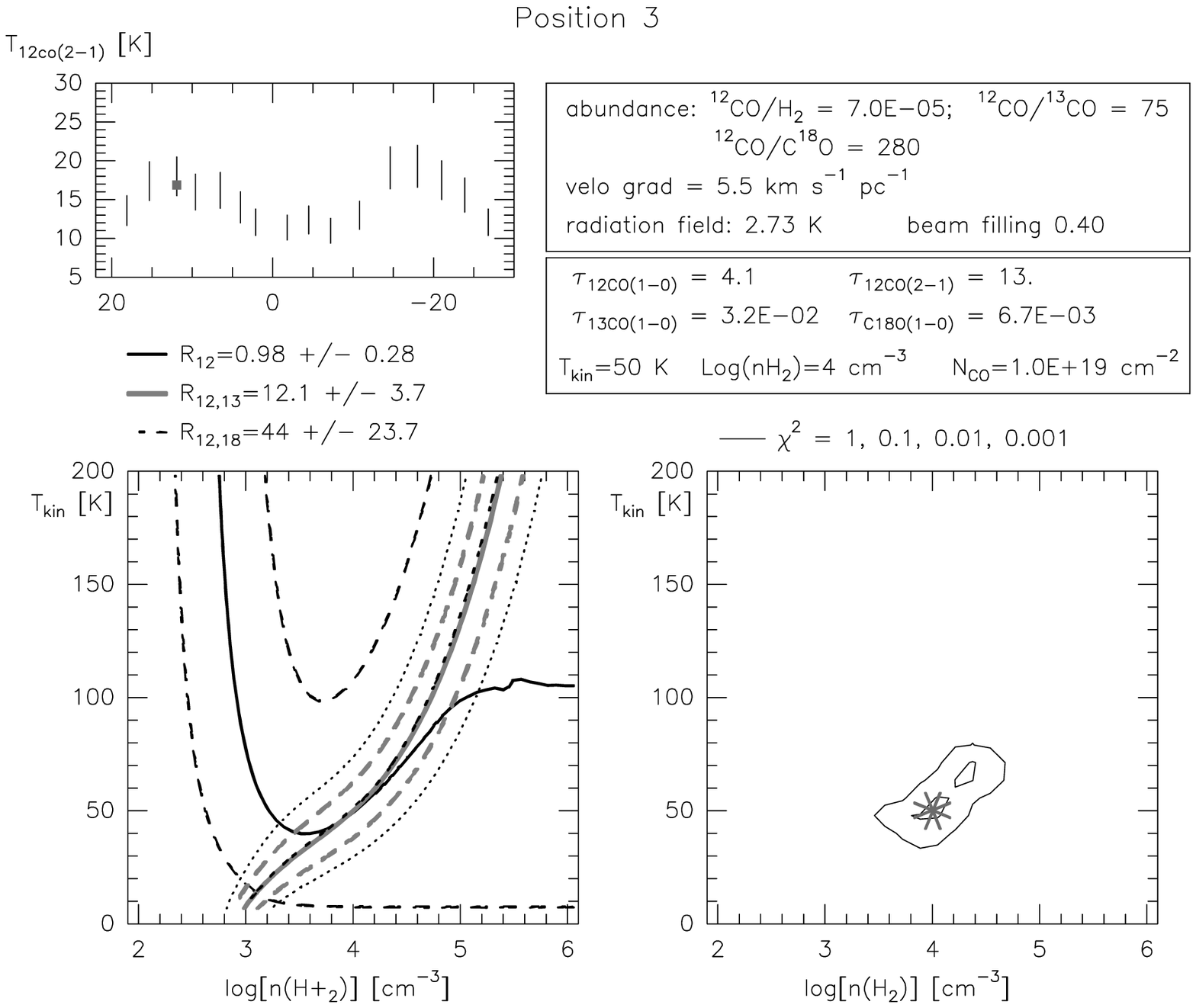}}
\caption{`Best' LVG solution at position 3 (the eastern CO lobe). The top left 
diagram shows the observed \bco\, line intensities across the major axis of 
M\,82 and the \bco\, intensity as given by the LVG--model for position 3. 
The fixed LVG--parameters (abundances, velocity gradient, radiation field and 
beam filling) are given in the upper right box. The parameters below the LVG 
input parameters summarize the optical depth for each transition, the CO column 
density, the kinetic temperature and \hh\, density for the `best fit'. The 
black and the grey solid line in the lower left diagram are the \bco/\aco\, 
(R$_{12}$) and \cco/\aco\,(R$_{12,13}$) line intensity ratio, respectively. 
The dashed lines indicate the observational errors for the 
corresponding line ratio. The \eco/\aco\,(R$_{12,18}$) line intensity 
ratios are given by the dashed--dotted line. Their errors are indicated by 
the dotted line. The contours in the 
lower right diagram are results of the $\chi^2$ test comparing the predicted 
line ratios and the \bco\, line intensity with the observed values. 
The star indicates the best solution in the kinetic temperature and \hh\, 
density plane.}
\label{lvg-pos3}
\end{figure*}
\end{center}

\begin{center}
\begin{figure*}
\resizebox{14.0cm}{!}{\includegraphics{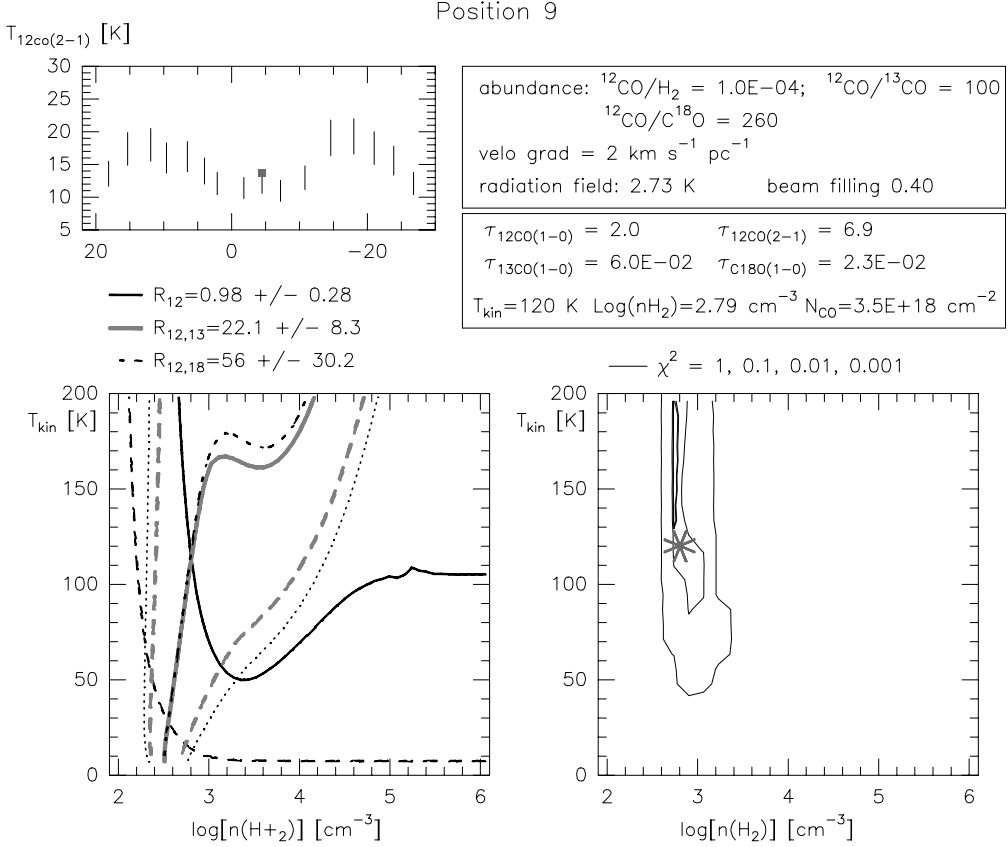}}
\caption{`Best' LVG solution at position 9 (central MIR peak). The items 
and parameters are the same as in Fig\,\ref{lvg-pos3}.}
\label{lvg-pos9}
\end{figure*}
\end{center}

\begin{figure*}
\begin{center}
\resizebox{12cm}{!}{\includegraphics{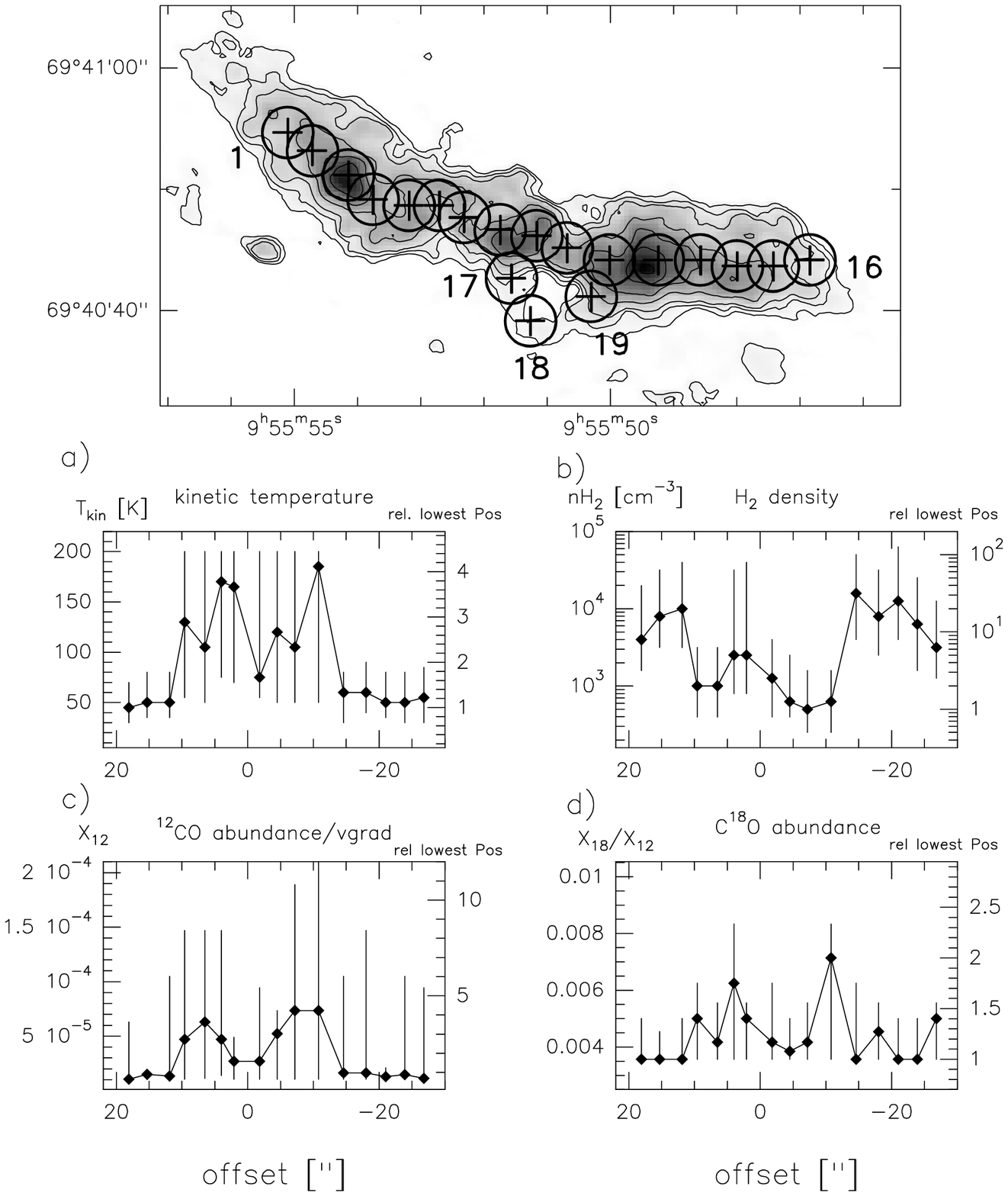}}
\caption{LVG solutions for positions 1 to 16. Top: locations of the analyzed 
positions. The radii of the circles indicate the spatial resolution for which 
the line ratios have been determined. a) to d): spatial variations of the kinetic 
temperature, the \hh\, density, the CO abundance per velocity gradient and the
fractional \abc\, abundance across the major axis of M\,82. The error bars in a) 
and b) correspond to the parameter range of kinetic temperatures and \hh\, 
densities for which the LVG line ratios and the \bco\, intensity is consistent 
with the observations within the errors. This corresponds to the area within the
$\chi^2=1$ contour shown for position 3 and 9 in Figs\,\ref{lvg-pos3} and 
\ref{lvg-pos9} . The error bars in c) and d) correspond to the range for 
[CO]/grad(V) and fractional \abc\, abundance where the $\chi^2$ of the 
corresponding fit is 50 times higher than for the `best' solution.}
\label{lvg-para}
\end{center}
\end{figure*}

The excitation conditions of the CO--emitting volume were modeled using a 
spherical, isothermal one--component large velocity gradient (LVG) model 
(Goldreich \& Kwan \cite{goldreich}, de Jong \etal\ \cite{dejong}). LVG line 
intensities were calculated for a kinetic temperature and \hh\, density range 
from 5\,K to 200\,K by 5\,K and ${\rm log}\,n(H_2)$  from 1.8 to 5.0 by 0.2 
respectively. In addition, we varied the CO abundance relative to \hh, [CO], 
per velocity gradient and the fractional \abb\, and \abc\, abundances 
([CO]/grad(V): $1\times10^{-5}\,{\rm to}\,2\times10^{-4}\,{\rm by}\,
1\times10^{-6}$; [CO]/[\abb]: 30 to 100 by 5; [CO]/[\abc]: 100 to 300 by 20). 
For the comparison between the observed peak intensity ratios 
(Table \ref{lineratios}, columns 3,5,7) and the predicted LVG ratios we used a 
${\chi}^2$ test. To account for the absolute intensities across the disk
of M\,82 we also fitted the \bco\, intensity at each position by 
varying the beam filling from 0.1 to 0.9 by 0.1. The `best' solutions are 
shown for positions 3 and 9 in Figs.\ \ref{lvg-pos3} \& \ref{lvg-pos9}. 
Position 3 on the western CO lobe is an example for a solution with low 
kinetic temperatures and high \hh\, densities; position 9 on the brightest
MIR peak is representative for solutions with high kinetic temperatures and 
low \hh\, densities.\\
The observed line ratios and \bco\, intensities can be modeled 
within the errors at all positions. The fit agrees very well with the data 
at positions where \bco/\aco\, is less than 1.2. At position 6 (eastern MIR peak) 
we do not find any intersection for all observed line ratios in the \hh\, 
density and kinetic temperature plane. For a more detailed discussion see 
Sect.\ \ref{diskussion}. The best agreement with the observed line ratios and 
absolute intensities is found for a beam filling of 0.4. Positions 6 and 7 
at the eastern MIR peak (Telesco \& Gezari \cite{telesco92}) and positions 
18 and 19 at the CO outflow require a somewhat lower beam filling of 0.2 and 
0.3 respectively. \\ 
The LVG parameters of the `best--fit' across the major axis of M\,82 are shown 
in Figs.\ref{lvg-para}\,a--d. The CO abundance relative to \hh\, per velocity 
gradient ([CO]/grad(V)) varies between $1\times10^{-5}$\,pc/\kms\, and 
$7\times10^{-5}$\,pc/\kms. Assuming ${\rm grad(V)}\approx 1 {\rm \kms\,pc^{-1}}$, 
as suggested by comparing the linewidth with the linear extent of the region, 
this corresponds to CO abundances in the range of ${\rm [CO]}\approx 
10^{-5}-7\times 10^{-5}$. Similar values have been determined 
in the Orion region (Blake \etal\ \cite{blake87}) and were suggested by 
chemical models  (Farquhar \etal\ \cite{farquhar}). [CO]/grad(V) 
increases towards the MIR peaks which indicates higher CO abundances at the 
active star--forming regions than in the more quiescent outer regions.
The fractional \abb\, abundance [\aba]/[\abb] across M\,82 does not show 
any significant spatial variation. The mean value of all positions is 
$70\pm20$. A low fractional \abb\, abundance is consistent with 
recent radiative transfer calculations by Mao \etal\ (\cite{mao}) and an 
independent chain of arguments based on CN and $^{13}$CN measurements 
(Henkel \etal\ \cite{henkel98}). In contrast, the fractional \abc\, 
abundance [\aba]/[\abc] shows a trend towards higher \abc\, abundances at 
the MIR peaks and in the outflow. While the average [\aba]/[\abc] ratio in 
the quiescent regions is about 270, it is only about 160 at position 6, 11,
17 and 19 (see Fig.\ \ref{lvg-para}\,d). Note that these values suggest \abc\, 
abundances 2--3 times higher than those used by Wild \etal\ (\cite{wild}) 
for their LVG calculations of CO line ratios in M\,82.\\ 
The kinetic temperature is well correlated with the MIR emission and other 
tracers of high--level star formation. Within the prominent CO lobes, with 
less signs of ongoing star formation, the kinetic temperature is about 50\,K.
Towards the active star--forming regions we find two kinetic temperature 
peaks above 150\,K. These `hot--spots' coincide with the location of MIR peaks 
(for a comparison between the MIR emission and the CO distribution see 
Telesco \& Gezari \cite{telesco92}). Near the 2.2\m\, nucleus the LVG models 
suggest temperatures of about 75\,K. Along the CO outflow the temperature 
drops with increasing distance from the active regions. At position 17 and 19 
we find temperatures above 100\,K. At position 18 (100\,pc distance from 
the plane) the kinetic temperature has dropped to 60\,K. The spatial variation 
of the kinetic temperature along the major axis of M\,82 is shown in 
Fig.\ \ref{lvg-para} a. The corresponding diagram of the \hh\, density 
distribution is shown in Fig.\ \ref{lvg-para} b. Solutions are found between 
$n(H_2)= 10^{2.7}$ and $10^{4.2} {\rm cm}^{-3}$. In general, the \hh\, densities 
are high in regions with low kinetic temperatures and vice versa. The solutions 
for the outer CO lobes suggest an \hh\, density about $n(H_2)=10^{4.0} 
{\rm cm}^{-3}$ with a tendency towards somewhat lower values at the very edge of 
the CO distribution ($n(H_2)=10^{3.5} {\rm cm}^{-3}$). These values 
are in agreement with \hh\, densities calculated by Wild \etal\ (\cite{wild}) 
and Mao \etal\ (\cite{mao}). At the `hot--spot', low \hh\, densities of 
$n(H_2)=10^{2.8-3.1} {\rm cm}^{-3}$ are required to match the observed line 
ratios. \hh\, densities in the CO outflow are about  $n(H_2)=10^{3.0} 
{\rm cm}^{-3}$. \\ 
Both the \aco\, and the \bco\, transitions are optically thick. 
In the cold dense regions we find an optical depth of  $\tau_{\aco} = 2-5$ and 
$\tau_{\bco} = 7-15$. At the `hot--spots' the derived optical depths are somewhat
lower and reach unity in the \aco\, transition at the eastern MIR peak 
(position 6 \& 7). For the ground transitions of the rare isotopes \abb\, and 
\abc\, we find optically thin emission at all positions. Typical 
optical depths are $\tau_{\cco} = 5\times10^{-2}$ and 
$\tau_{\eco} = 5\times10^{-3}$.

\subsection{Column densities and mass distribution}

\begin{figure}[b]
\resizebox{8.5cm}{!}{\includegraphics{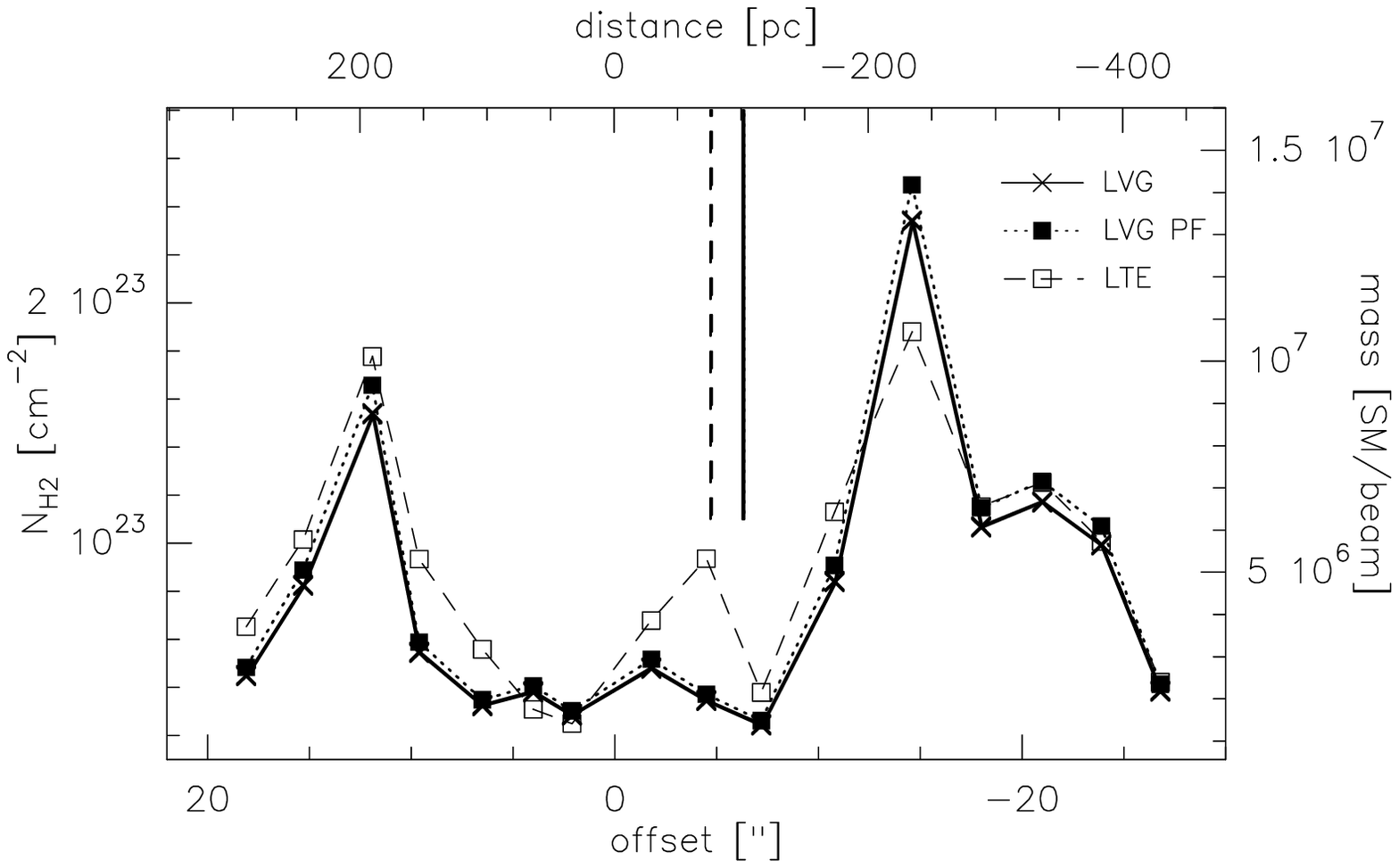}}
\caption{Beam averaged \hh\,column densities across the major axis of M\,82. 
Offsets are given relative to the centre of M\,82 (2.2\m\, peak). The thick 
solid line corresponds to the \hh\, column densities derived from the LVG 
solutions. The dashed and the dotted lines give the corresponding values for 
the LTE approximation and the solution calculated from the partition functions 
(LVG$_{PF}$), respectively. The vertical lines indicate the location of the 
centroid of mass of the three \hh\, distributions. Note that the location 
of the centroid of mass is identical for the LVG and LVG$_{PF}$ solution}. 
The labeling of the right axes gives the \hh\, mass per beam in units of 
solar masses.\label{columndensity}
\end{figure}

For the determination of  CO and \hh\, column densities at each position we 
used three methods:\\
-- LVG: The column densities were derived from the CO and \hh\, densities, 
the velocity gradient and the observed line widths using ${\rm N(CO)}=3.08\times 
10^{18}\,n({\rm CO})\,\frac{dV}{{\rm grad(V)}}$ and ${\rm N(\hh)}=
3.08\times 10^{18}\,n(\hh)\,\frac{dV}{{\rm grad(V)}}$, where $dV$ is the 
observed line width. Therefore $\frac{dV}{{\rm grad(V)}}$ is an equivalent 
path length through the clouds.\\
-- LVG$_{PF}$ (PF\,=\,partition function): The \abb\, and \abc\, column 
densities were calculated from the general relation between optical depth, 
excitation temperature and column density at rotation level $J$:
$N_J=93.5\frac{g_J \nu^3}{g_{(J+1)} A_{J+1,J}}(1-{\rm exp}(-4.8\,10^{-2} 
\nu/T_{ex}))^{-1} \int \tau dv$ where $g_J$ is the statistical weight of 
level $J$ and $A_{J+1,J}$ is the Einstein coefficient for the 
transition $J+1$ to $J$. $\int \tau dv$ was approximated by $\int \tau 
dv \approx 1.06\, \tau \,dV$. $T_{ex}$, $\tau$, [\abb] and [\abc] 
are given by the LVG code for each level.  \abb\, and \abc\, column densities 
were determined using the sum of the 6 lowest levels for each isotope. 
\hh\, and CO column densities were derived from the relative abundances 
of the rare isotopes relative to \hh\, and CO.\\
-- LTE: \abb\, and \abc\, column densities were derived using a standard LTE 
approach (e.g. Dickman \cite{dickman}). As for the LVG$_{PF}$ method, CO and 
\hh\, column densities were derived from the abundances of the rare 
isotopes relative to CO and \hh at each position.\\
Column densities calculated from \abb\, and \abc\ via the LTE method match 
each other with less than 5\% difference at each position. The same holds
for the LVG$_{PF}$ column densities calculated from $T_{ex}$ and $\tau$ of 
the \abb\, and \abc\ transition. For simplicity we therefore give in the 
following the average between the column densities calculated from \abb\, 
and \abc\ via the LTE and LVG$_{PF}$ method.\\ 
The spatial variations of the beam--averaged \hh\, column density across the 
major axis of M\,82 as calculated with the three methods is shown in 
Fig.\ \ref{columndensity}. The spatial distribution of the \hh\, column densities 
is in good agreement for all three methods. This suggests that the low $J$ levels 
are almost thermalized. The largest difference between the methods is apparent at 
the central CO peak. While the LTE solutions suggest 
a local \hh\, column density maximum of about ${\rm N(\hh)_{4''}}=1\times10^{23}
\,{\rm cm^{-2}}$, the peak is less prominent (${\rm N(\hh)_{4''}}=5\times10^{22}
\,{\rm cm^{-2}}$) and displaced by $4''$ in the LVG and LVG$_{PF}$ solution 
(see Fig.\ \ref{columndensity}).

\noindent Nevertheless, all methods clearly show that most of the molecular 
gas traced by CO is located in the outer CO lobes. The central 300\,pc between 
the molecular lobes contain only about 20--30\% of the molecular gas mass. 
Furthermore, the \hh\, column density distribution is clearly asymmetric with 
respect to the 2.2\m\, nucleus. We find that the centroid of mass is located 
about 100\,pc south--east of the nucleus. The location of the centroid of 
mass for each method is indicated by the vertical line in Fig. \ref{columndensity}.
The highest \hh\, column density is found at the western CO lobe (position 12). 
Its beam--averaged LVG column densities are ${\rm N(CO)_{4''}}=2\times10^{19}\,
{\rm cm^{-2}}$ and ${\rm N(\hh)_{4''}}=2.3\times10^{23}\,{\rm cm^{-2}}$. 
The corresponding cloud--averaged LVG column densities are ${\rm N(CO)_{cloud}}=
4\times10^{19}\,{\rm cm^{-2}}$ and ${\rm N(\hh)_{cloud}}=6\times10^{23}\,
{\rm cm^{-2}}$, respectively. The corresponding values for the eastern
CO lobe (position 3) are ${\rm N(CO)_{4''}}=1\times10^{19}\,{\rm cm^{-2}}$, 
${\rm N(\hh)_{4''}}=1.5\times10^{23}\,{\rm cm^{-2}}$, ${\rm N(CO)_{cloud}}=
2.5\times10^{19}\,{\rm cm^{-2}}$ and ${\rm N(\hh)_{cloud}}=3.8\times10^{23}\,
{\rm cm^{-2}}$. For an assumed line--of--sight of 350\,pc (for comparison with 
Mao \etal\ \cite{mao}) the mean molecular density in the CO lobes is 
$<n(\hh)>_{4''}= 140 - 210\,{\rm cm}^{-3}$. This corresponds to a volume filling 
factor of $f_{v,4''}=<n(\hh)>_{4''}/n(\hh)\approx 0.01$. With $f_{a,4''}=0.4$ 
and a linear resolution of 65\,pc we obtain characteristic cloud sizes of 
$r_{cloud}= \frac{1}{2}\, 65\,{\rm pc}\,f_{v,4''}/f_{a,4''} \approx 1{\rm pc}$. 
Volume filling factors and characteristic cloud sizes do not change 
significantly in the central star forming regions. These values are in 
good agreement with PDR models published 
by Wolfire \etal\ (\cite{wolfire}).\\ \hh\, column densities in the molecular 
outflow are in the range ${\rm N(\hh)_{4''}}=1.5-3.0\times10^{22}\,{\rm cm^{-2}}$. 
The total mass of the outflow is $7.2 \times 10^5\,\msol$ (D\,=\,3.9 Mpc, 
Sakai \& Madore \cite{sakai}).\\

\subsection{Conversion from I(CO) to \nhh\, and total mass}

\begin{figure}[t]
\resizebox{8.5cm}{!}{\includegraphics{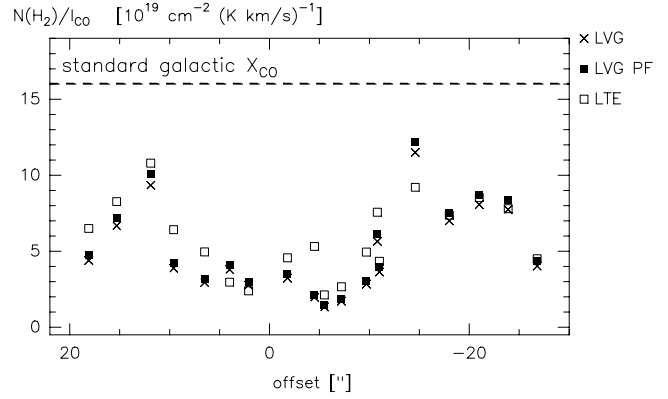}}
\caption{Variation of \xco\, across the analyzed positions in M\,82. Offsets are 
given relative to the centre of M\,82 (2.2\m\, peak). The \hh\, column densities 
were derived using the LTE, the LVG and LVG$_{PF}$ solutions. 
The dotted line at $\xco\,=1.6\times10^{20}\,{\rm cm^{-2} \,(K \kms)^{-1}}$ 
corresponds to the standard Galactic conversion factor.}
\label{xco}
\end{figure}

\begin{figure}[t]
\resizebox{8.5cm}{!}{\includegraphics{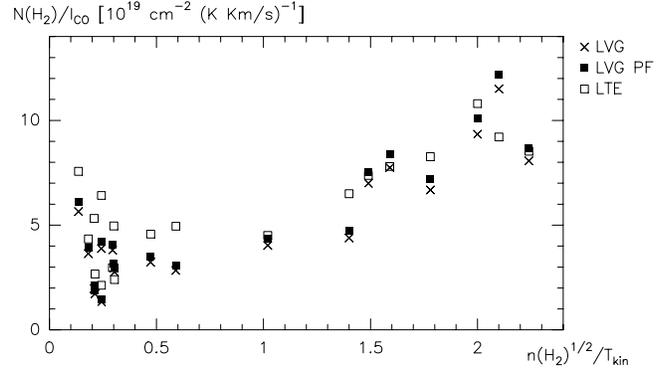}}
\caption{\xco\, as derived from the LTE, LVG and LVG$_{PF}$ solutions 
versus ${\rm T}^{-1}\,{\rm n(\hh)}^{1/2}$. The plot shows a linear 
correlation between the conversion factor and the associated \hh\, densities 
and kinetic temperatures beyond ${\rm T}^{-1}\,{\rm n(\hh)}^{1/2}\approx 0.5$.}
\label{xco-var}
\end{figure}

\begin{figure}[h]
\resizebox{8.5cm}{!}{\includegraphics{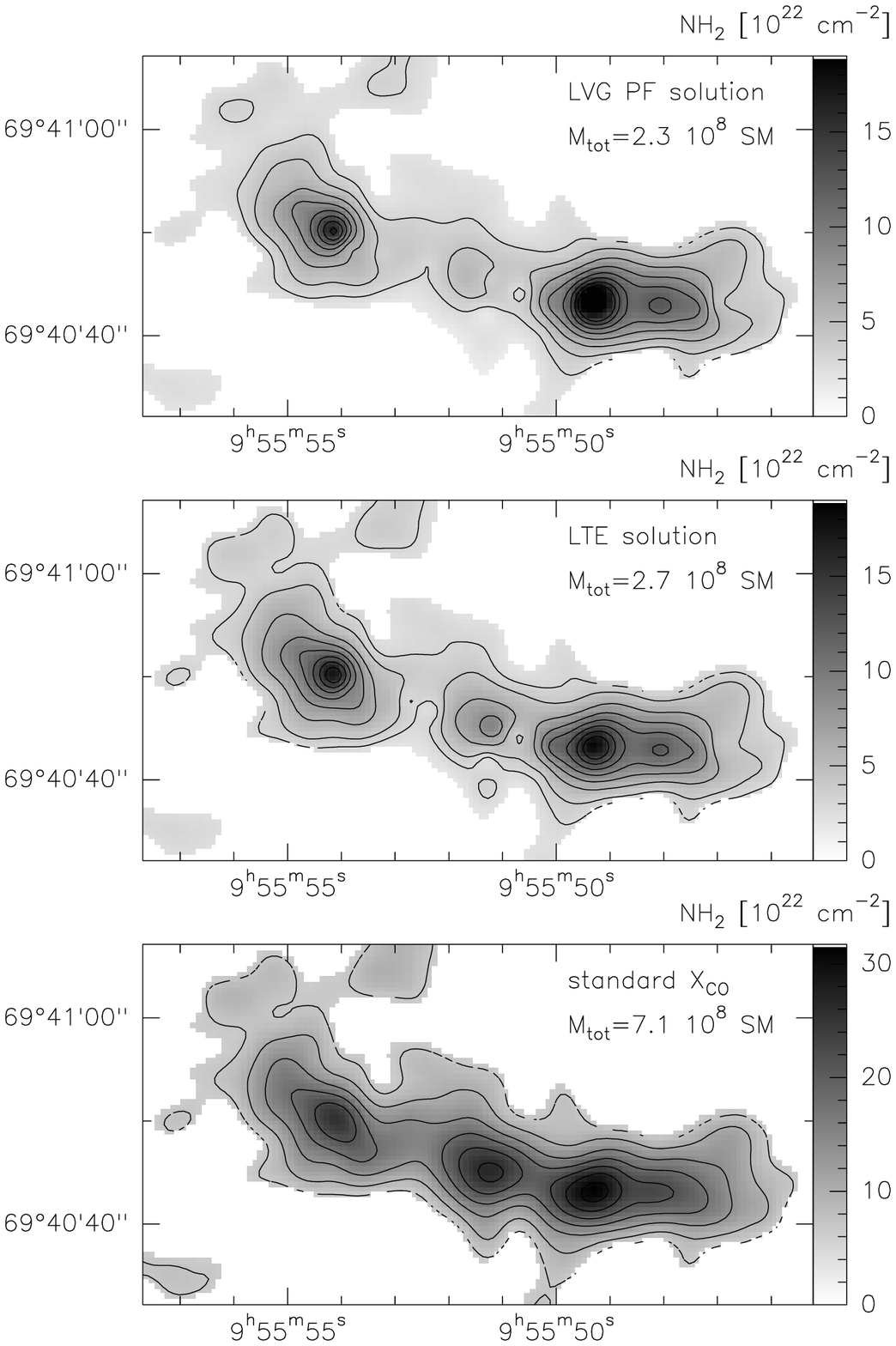}}
\caption{\hh\, column density maps calculated from the  integrated \aco\, 
intensity distribution at $4.2''$ resolution. The \hh\, maps were derived 
using the \xco\, distribution across the major axis of M\,82 as calculated 
from the LVG$_{PF}$ solutions (top), from the LTE solutions (middle) and from
the standard Galactic \xco\, conversion factor ($1.6\times10^{20}\,
{\rm cm^{-2} \,(K \kms)^{-1}}$) (bottom).  The total \hh\, mass is denoted in 
the upper right corner of each plot. Contour levels correspond to
1.4, 2, 3, 4, 6, 8, 10, 12, 14 and $16\times10^{22}\,{\rm cm^{-2}}$ (top),
3, 4, 6, 8, 10, 12, 14 and $16\times10^{22}\,{\rm cm^{-2}}$ (middle) and 
7, 10, 14, 18, 22, 26 and $30\times10^{22}\,{\rm cm^{-2}}$ (bottom).}
\label{nh2-maps}
\end{figure}

To derive the conversion factor from I(CO) to \nhh, we have compared LVG, LTE, and 
LVG$_{PF}$ \hh\, column densities with the integrated \aco\, intensities at 
$4.2''$ resolution at the analyzed positions across the central part of M\,82. 
The variation of the conversion factor $\xco\,=\nhh/{\rm I(CO)}$ with position 
is shown in Fig.\ \ref{xco}. Note that \xco\, is lower than the Galactic value 
of $1.6\times10^{20}\,{\rm cm^{-2}\,(K \kms)^{-1}}$ (Hunter \etal\ 
\cite{hunter97}) at all positions and for all methods. We find that \xco\, 
varies across the disk of M\,82 by about a factor of 5 if one considers the 
LTE solutions ($\xco\,=2.1-10.8\times10^{19}\,{\rm cm^{-2} \,(K \kms)^{-1}}$) 
and by a factor of 8--9 for the LVG and LVG$_{PF}$ solutions 
($\xco\,=1.3-11.5\times10^{19}
\,{\rm cm^{-2} \,(K \kms)^{-1}}$ and $\xco\,=1.5-12.2\times10^{19}\,
{\rm cm^{-2}\,(K \kms)^{-1}}$) . All methods show that the lowest conversion 
factors are associated with the central star--forming regions where the 
gas is heated by UV photons from the newly formed stars and cosmic--rays from SNRs. 
The CO--emitting volumes at these positions have high kinetic temperatures. 
Towards the outer molecular lobes with higher \hh\, densities and lower kinetic 
temperatures, the conversion factor rises. This is in agreement with simple 
theoretical arguments that suggest that the conversion factor \xco\, should 
be proportional to ${\rm T}_{\rm kin}^{-1}\,{\rm n(\hh)}^{1/2}$ for virialized 
clouds (Maloney \& Black \cite{maloney}). The variation of \xco\, with 
${\rm T}_{\rm kin}^{-1}\,\,{\rm n(\hh)}^{1/2}$ is shown in Fig.\ \ref{xco-var}. 
The linear correlation between  \xco\, and ${\rm T}_{\rm kin}^{-1}\,
{\rm n(\hh)}^{1/2}$ for ${\rm T}_{\rm kin}^{-1}\,{\rm n(\hh)}^{1/2}>0.5$ is 
clearly visible. For ${\rm T}_{\rm kin}^{-1}\,{\rm n(\hh)}^{1/2}<0.5$ the scatter 
in the plot increases. This is in particular true for \xco\, calculated under the 
assumption of LTE. This suggests that the gas is not close to LTE at the 
`hot spots'. The increased scatter of \xco\, calculated with the LVG and 
LVG$_{PF}$ method might suggest that either the clouds are not virialized 
or that more appropriate models (like PDR models) are required to calculate 
the physical gas properties in the centre of the starburst. For a more 
detailed discussion see Sect. \ref{diskussion}. Nevertheless, this 
result not only shows that the standard Galactic \xco\, factor is not 
appropriate for a starburst system like M\,82, but that {\it \xco\, is a 
function of the intrinsic gas properties which strongly depend on 
environmental effects}. This implies that spatial variations of \aco\, 
intensities can be due to variations of the excitation conditions of the gas 
rather than variations of column density. Similar results have been obtained by 
Wild \etal\ (\cite{wild}) using low--resolution CO data (see also Sect.\ 
\ref{dis-xco}). Based on the analysis of \xco\, we have calculated the `true' 
\hh\, distribution in M\,82 by interpolating the changes of \xco\, from the 
analyzed positions across the central CO distribution. Multiplication of this 
\xco\,--map with the integrated \aco\, intensity distribution thus results in an 
\hh\, column density map. We show these maps in Fig.\ \ref{nh2-maps} for \xco\, 
derived from the LVG$_{PF}$ (top) and LTE solutions (middle) in comparison with 
the \hh\, distribution one would derive assuming a constant, standard Galactic 
conversion (bottom) to illustrate the importance of detailed studies of \xco\, 
to derive \hh\, column density distributions. The \hh\, column density maps 
in Fig.\ \ref{nh2-maps} (top \& middle) indicate that the central star--forming 
region is surrounded by a double--lobed distribution of molecular gas, while 
\hh\ seems to be depleted in the central starburst region itself (see also Fig.\ 
\ref{columndensity}). \\ 
The total \hh\, mass of the region shown in Fig.\ \ref{nh2-maps} is 
$2.3\times 10^8\,\msol$ for the LVG$_{PF}$ and LVG and $2.7\times 
10^8\,\msol$ for the LTE solution at a distance of D\,=\,3.9 Mpc 
(Sakai \& Madore \cite{sakai}). The corresponding values at 
D\,=\,3.25 Mpc (Tammann \& Sandage \cite{tammann}) are $1.6$ and $1.9\times 10^8\,\msol$, 
respectively. These values are in good agreement with estimates from 450 \m\, dust 
continuum measurements (Smith \etal\ \cite{smith}) and from C$^{18}$O(2$\to$1) intensities 
(Wild \etal\ \cite{wild}). Therefore, the total molecular mass is 3 times lower 
than the mass one would derive using the standard Galactic conversion factor of 
$1.6\times10^{20}\,{\rm cm^{-2} \,(K \kms)^{-1}}$ ($4.9\times 10^8\,\msol$ D\,=\,3.25 Mpc; 
$7.1\times 10^8\,\msol$ D\,=\,3.9 Mpc).

\section{Discussion}

\subsection{Comparison with other radiative transfer analyses \label{diskussion}}

Analyses of the physical conditions of the molecular gas in M\,82 have been 
published by Tilanus \etal\ (\cite{tilanus}), Wild \etal\ (\cite{wild}), 
G\"usten \etal\ (\cite{guesten}) and more recently by Mao \etal\ (\cite{mao}) 
and Petitpas \& Wilson (\cite{petitpas}) using single--dish CO data and other 
tracers of the molecular gas. The kinetic gas temperature of the CO--emitting 
gas phase derived in these studies are 
of order ${\rm T}_{kin}=30-70\,{\rm K}$. \hh\, densities range between 
$n(\hh) = 10^3-10^4\,{\rm cm}^{-3}$. Thus our solutions at the CO lobes 
and the outer parts of the CO distribution ($n(\hh) = 10^{3.5-4.2}\,{\rm cm}^{-3}, 
{\rm T}_{kin}=45-60\,{\rm K}$) are consistent with previous studies. \\
The situation is different for our LVG solutions within the starburst region. 
Kinetic gas temperatures above $150\,{\rm K}$ are clearly inconsistent with 
results published in literature so far. In the most recent analysis of the 
excitation conditions of the molecular gas using mm and sub--mm CO emission 
lines (up to $J=7-6$), Mao \etal\ (\cite{mao}) suggested kinetic gas temperatures 
as high as 130\,K. However, they rejected their LVG solution because of intrinsic 
inconsistencies regarding almost equal area and volume filling factors derived from 
the one--component LVG model. This leads to characteristic cloud sizes of 
about 150\,pc which is inconsistent with high--resolution studies of the CO 
distribution in M\,82 (Shen \& Lo \cite{shen}, Neininger \etal\ \cite{nico} and 
this work). Characteristic cloud sizes derived from our analysis are only 
$\approx 1\,{\rm pc}$, hence more realistic. This difference results 
mainly from a very low area filling factor of only $f_{a,22''}=0.04-0.07$ found 
by Mao \etal\ (\cite{mao}). From the CO morphology 
(see Fig. \ref{m0-12co21-ssc}) and 
$f_{a,4''}=0.4-0.2$ we would expect area filling factors of $f_{a,22''}=0.2-0.15$ 
at $22''$ resolution. The reason for this discrepancy remains unclear. 
But obviously the assumption of an isothermal gas phase used in the LVG model 
is more reasonable for our high spatial resolution study than for the 
low--resolution data used by Mao \etal\ (\cite{mao}). Even though our 
LVG analysis does not 
lead to internal inconsistencies we also find that the observed line ratios 
are difficult to reproduce with the one--component LVG model at the `hot spots'. 
This is in particular true for positions 6, 7 \& 11 (eastern MIR peak and 
expanding superbubble) where no intersection of all observed line ratios 
(disregarding the errors of the observations) exists within the 
calculated parameter space. At these positions more sophisticated radiative 
transfer models like PDR models probably lead to more consistent results. However,
a comparison between LVG and PDR models in M\,82 shows that constraints on 
\hh\, densities and beam--averaged column densities are very similar for both 
methods (Mao \etal\ \cite{mao}). G\"usten \etal\ (\cite{guesten}) and 
Mao \etal\ (\cite{mao}) concluded 
that in order to explain the observed line ratios, a two--component model of 
the molecular gas in M\,82 is needed. However, densities of $n(\hh) = 10^{3.7}\,
{\rm cm}^{-3}$ for the high density and $n(\hh) = 10^{3.0}\,{\rm cm}^{-3}$ for 
the low density component (Mao \etal\ \cite{mao}) are much lower than 
values given by G\"usten \etal\ (\cite{guesten}) 
($10^{3.0}\,{\rm cm}^{-3}$, $10^{5.0}\,{\rm cm}^{-3}$). In addition the
high density component in the models by G\"usten \etal\ is associated with a high 
kinetic temperature while it is cooler than is low density component in the 
models by  Mao \etal\ (\cite{mao}). Our high--resolution LVG results 
confirm the PDR calculations of Mao \etal\ (\cite{mao}).
In particular, their conclusion that the bulk of CO emission in the core of 
M\,82 arises from a warm, low--density interclump medium is consistent with our 
findings. \\  
Further support for a highly excited CO component towards the active regions 
in M\,82 comes from the morphology of the high--J CO lines observed by 
Mao \etal\ (\cite{mao}). They find that the spatial separation of the CO 
lobes decreases with increasing J. While the spatial separation of the outer 
CO lobes in the \aco\ and \bco\ transition is about $26''$, it decreases 
to only $15''$ in the $^{12}{\rm CO}(J\!=\!7\to6)$ transition. This distance 
is in good agreement with the spatial separation of the kinetic temperature peaks 
that we find in our LVG solutions (Fig\,\ref{lvg-para}\,a).\\ 
To further test the reliability of the modeled kinetic temperature and \hh\, 
density distribution across M\,82 we calculated line ratios for the high--J 
transitions of \aba\, and \abb\, at $22''$ resolution and compared our prediction 
with the line ratios published by Mao \etal\ (\cite{mao}). Note that the spatial 
smoothing of our high--resolution one--component excitation model leads to a 
multi--component model at lower resolution because it encompasses the individual 
solutions (weighted with a Gaussian of $22''$ width) at all positions. 
The predicted line ratios from a single CO isotope match the observations 
extremely well. For different CO isotopes the predicted line ratios are lower 
than suggested by the observations, but consistent within the observational errors. 
The predicted and observed line ratios are summarized in Table\,{\ref{high-ratios}.\\
An independent chain of arguments favoring a temperature gradient towards the central 
starburst region comes from the different separation of the eastern and western 
`hot spots' in MIR and FIR observations. Hughes \etal\ (\cite{hughes}) stated 
that the larger separation of the peaks at 450 \m\, reflects the radial 
temperature gradient that must exist within the torus if the dust is heated 
by the central starburst population.

\begin{table}
\caption{Line ratios for the high--J transition of \aba\, and \abb\, at the 
centre and the eastern and western CO lobe of M\,82. The line ratios are given 
for a spatial resolution of $22''$. The first row of each line ratio gives the 
value calculated from the LVG model. The second and third row are the observed 
peak and integrated line ratios adopted from Mao \etal\ (\cite{mao}).
\label{high-ratios}}
\begin{tabular}{l c c c l}
CO line ratio & east & centre & west &\\ \hline
$^{12}$CO(7$\to$6)/$^{12}$CO(4$\to$3) & 0.36 & 0.25 & 0.31 &model\\
                                & 0.32 & 0.29 & 0.37 & peak \\
                                & 0.28 & 0.36 & 0.33 & integral\\
& & & & \\
$^{12}$CO(4$\to$3)/$^{13}$CO(3$\to$2) & 7.07 & 8.02 & 5.81&model \\
                                & 8.50 & 10.1 & 9.07 & peak \\
                                & 9.09 & 6.88 & 7.40 & integral\\
& & & & \\
$^{13}$CO(2$\to$1)/$^{13}$CO(1$\to$0) & 1.92 & 1.67 & 1.87&model \\
                                & 1.53 & 1.63 & 1.93 & peak \\
                                & 1.47 & 1.44 & 1.17 & integral\\
& & & & \\
$^{12}$CO(2$\to$1)/$^{13}$CO(2$\to$1) & 7.82 & 9.27 & 7.48&model \\
                                & 9.04 & 13.7 & 9.19 & peak \\
                                & 11.4 & 14.2 & 11.7 & integral\\
\end{tabular}
\end{table}

\subsection{The state of the molecular gas in M\,82's centre}
 
Our study does not imply that the central part of M\,82 does not contain dense
molecular cloud cores as traced by other molecular species like HCN (see e.g. 
Seaquist \etal\ \cite{seaquist}; Brouillet \& Schilke \cite{brouillet}). Most of 
the CO emission from the core of M\,82, however, arises from a warm, low density 
interclump medium. This conclusion was already reached by Mao \etal\ 
(\cite{mao}) using PDR 
models. The existence of such a gas component, which we can be confident of, raises 
the question if it can survive under the extreme conditions in the starburst 
centre or whether it is indicative of cloud evaporation and thus 
of a destruction of molecular clouds. From the observed UV field strength one would 
expect that the molecular clouds with densities less than $10^4\,{\rm cm}^{-3}$ 
cannot exist in M\,82's centre (Brouillet \& Schilke \cite{brouillet}). 
In addition, such a gas component is barely dense enough to avoid tidal disruption 
(Mao \etal\ \cite{mao}) which would support a rather diffuse 
distribution of the gas. Such a scenario would explain the high [C\small I}]/[CO] 
abundance ratios observed in M\,82 (Schilke \etal\ \cite{schilke}; 
White \etal\ \cite{white}) and the depletion of \hh\ in the central 300\,pc 
of M\,82. As a consequence, this would imply that the clouds are not virialized. 
This could explain why the linear correlation between ${\rm T}_{\rm kin}^{-1}\,
{\rm n(\hh)}^{1/2}$ and \xco\, is not valid for ${\rm T}_{\rm kin}^{-1}\,
{\rm n(\hh)}^{1/2}<0.5$, hence for the warm and low--density gas in the starburst 
centre (see Fig.\ \ref{xco-var}). Therefore, there seems to be observational 
evidence that molecular clouds are partly disrupted and dissociated in the 
starburst centre. More detailed numerical analysis of evaporation
time scales and related questions are needed to settle this question.      

\subsection{\xco\, \label{dis-xco}}    

The suitability of a global Galactic factor \xco\, to convert \aco\, intensities 
to \hh\, column densities has been discussed by many authors (e.g. Young \& 
Scoville \cite{nic}, Bloemen \etal\ \cite{bloemen}, Hunter \etal\ 
\cite{hunter97}). As seen, theoretical studies of \xco\, showed that it is 
sensitive to the kinetic temperature of the emitting gas and that the 
conversion factor should be lower for starburst galaxies like M\,82 than for the 
Milky Way (Maloney \& Black \cite{maloney}). Investigations of \xco\, in M\,82 
confirmed this prediction: Wild \etal\ (\cite{wild}) used the optically thin 
C$^{18}$O(2$\to$1) transition to derive \hh\, column densities and hence 
\xco\, along the major axis of M\,82. They found $\xco\, \leq\,1 \times10^{20}
{\rm cm^{-2}\,(K \kms)^{-1}} $ and variations by a factor of 2 along the major 
axis. Similar results were obtained by Smith \etal\ (\cite{smith}) using
the 450 \m\, continuum emission from dust grains to derive \hh\,column densities. 
Even though both studies suggest a low conversion factor, its variation across 
the major axis shows significant differences. While Smith \etal\ (\cite{smith}) 
found that 
\xco\, decreases from east to west, with no changes in the central star--forming 
regions, the results by Wild \etal\ (\cite{wild}) suggest very 
low conversion factors near the 
eastern MIR peak and an increasing \xco\, towards the western molecular lobe. 
Therefore our variations of \xco\, in general support the results by Wild 
\etal\ (\cite{wild}). Nevertheless our conversion factors are slightly 
lower than those inferred by Wild \etal\ (\cite{wild}), and the location of 
the western \xco\, maximum 
is displaced by $\approx\,7''$. The discrepancies between the \xco\, factors 
derived from the molecular lines (Wild \etal\ \cite{wild} and this work) and the 
estimates from the 
dust emission might result from the single--temperature model used by 
Smith \etal\ (\cite{smith}). In particular in the central region, which shows 
strong MIR emission from heated dust (Telesco \& Gezari \cite{telesco92}), this might 
lead to an overestimate of the \hh\, column density and thus to an overestimate 
of \xco\,. Furthermore, 
the different morphology of the 450 \m\, map published by Hughes \etal\ 
(\cite{hughes}) raises doubts on the reliability of the 450 \m\, intensities 
used by Smith \etal\ (\cite{smith}) for their calculation. From this we 
conclude that the \xco\
factor in M\,82 is not only lower than the standard Galactic conversion factor, 
but that in addition \xco\ in the central 300\,pc is at least 3 times lower than 
in the molecular lobes. A similar gradient for the conversion factor has been 
found in the Milky Way towards the Galactic Centre  (e.g. Blitz \etal\ \cite{leo}, 
Sodroski \etal\ \cite{sodroski}, Dahmen \etal\ \cite{dahmen}). Furthermore, our 
analysis suggests that the variations of \xco\, are mainly caused by variations 
of the kinetic temperature of the CO--emitting volume due to environmental effects 
while abundance variations play a minor role.

\section{Conclusions}

We have observed the \bco\, and \eco\, emission lines in the starburst galaxy 
M\,82 with high spatial resolution using the Plateau de Bure interferometer. Our
main conclusions are:\\
\\
1) The overall morphology and kinematics for both transitions are similar to those of 
\aco\, and \cco\, published by Shen \& Lo (\cite{shen}) and Neininger \etal\ 
(\cite{nico}). The dynamical centre of the molecular gas coincides with the 
2.2 \m\, nucleus while the centroid of the molecular mass is located 100\,pc 
west of M\,82`s centre. South of the expanding molecular superbubble (Wei{\ss} 
\etal\ \cite{aweiss}) an outflow of molecular gas with a total mass of 
$7.2 \times 10^5\,\msol$ is detected.\\
2) The \bco/\aco\, line intensity ratios are lower ($\leq\,1.4$) than previously 
reported. Thus, CO line ratios in M\,82 are not outstanding, but comparable .
to values found in other starburst galaxies like NGC\,253. Line ratios vary 
across the disk of M\,82. Near the MIR peaks, the \bco/\aco\, ratios are high; 
in the outer parts, that are less affected by the starburst, this ratio drops to 
unity.\\
3) An LVG excitation analysis of the CO lines suggests that the excitation 
conditions of the molecular gas are strongly influenced by environmental 
effects. In the outer parts of the CO distribution we find \hh\, densities of 
$n(\hh) = 10^{3.5-4.2} {\rm cm}^{-3}$ and kinetic temperatures of 
${\rm T}_{kin}=45-60{\rm K}$. Towards the star--forming regions, indicated by 
strong MIR emission, the kinetic temperatures raise above 150\,K. The hot gas 
is associated with low \hh\, densities of only $n(\hh) = 10^{2.8-3.1} 
{\rm cm}^{-3}$. Area filling factors of $f_a = 0.2 - 0.4$ and volume filling 
factors of $f_v = 0.001 - 0.02$ indicate that the gas is organized in small 
clumps with a typical size of $r_{cloud} \approx 1\,{\rm pc}$. [\aba]/[\abb] 
abundance ratios are about 70 without significant spatial variations across 
the galaxy. In contrast, [\aba]/[\abc] abundance ratios in the outer parts of 
M\,82 are comparable to those found at the Galactic Centre ([\aba]/[\abc] = 270) 
but decrease to only [\aba]/[\abc] = 160 at the star--forming regions. 
Beam--averaged \hh\, column densities range from ${\rm N(\hh)_{4''}}=2.4
\times10^{22}\,{\rm cm^{-2}}$ near the MIR peaks to ${\rm N(\hh)_{4''}}= 
2.3\times10^{23}\,{\rm cm^{-2}}$ at the western CO lobe. The \hh\, 
distribution has a double--peak morphology which surrounds the central 
starburst region. The central 300\,pc are depleted in \hh. Thus the \hh\, 
distribution differs from the CO distribution. This result even holds when 
the \hh\, column densities are calculated under the assumption of LTE conditions.
The total molecular mass is $2.3 \times 10^8\,\msol$.\\
4) The conversion factor from $I$(CO) to \nhh\, ($X_{\rm CO}$) depends on 
the excitation conditions of the CO--emitting volume. Even in regions which are 
less affected by the starburst \xco\, is about 3 times less than the standard 
Galactic value. From the LVG analysis we find that  $X_{\rm CO} \sim 
{\rm T}_{\rm kin}^{-1}\,\,{\rm n(\hh)}^{1/2}$. Therefore  \xco\, is lower in 
the central star--forming regions than in the outer molecular lobes.

\section*{Appendix A: Short--Spacing Correction \label{app}}

\begin{figure}
\resizebox{9.0cm}{!}{\includegraphics{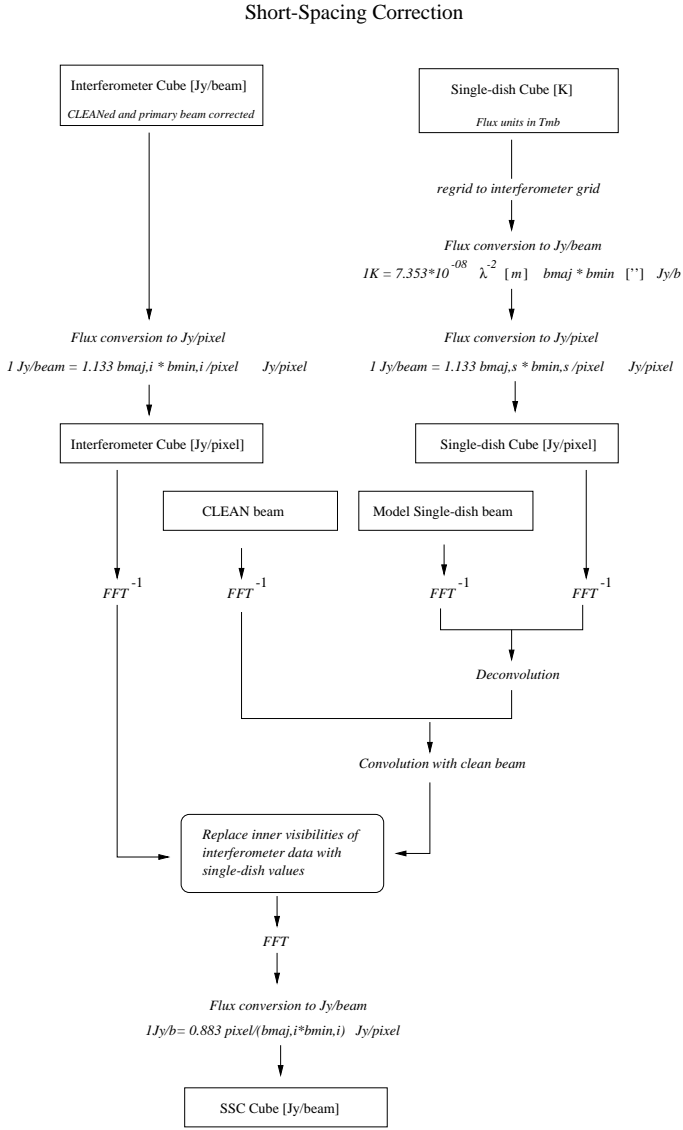}}
\caption{Flow chart of the short--spacing correction.}
\label{ssc-dia}
\end{figure}

A major difficulty related to flux determinations from interferometer maps is the 
missing flux problem. In brief it arises from the lack of coverage at low 
spatial frequencies in interferometric observations. This leads to an 
insensitivity to emission more extended than $\lambda/S^{proj}_{min}$, 
where $S^{proj}_{min}$ is the length of the shortest projected baseline. 
The recovered flux therefore only represents clumpy parts of a brightness 
distribution which leads to an underestimate of the total flux of a source. 
For a more detailed description see e.g. Wilner \& Welch \cite{wilner} and 
Helfer \& Blitz \cite{helfer}. The only way to overcome this problem is to derive 
the visibility at the origin of the {\it uv}--plane (which represents the 
integrated flux of a brightness distribution) and the low spatial frequencies 
from single--dish measurements and combine them with the interferometric 
observations. Methods for the combination of single--dish and interferometer data
have been described e.g. by Vogel \etal\ (\cite{vogel}) and Herbstmeier \etal\ 
(\cite{herbst}). Both methods generate the central visibilities from the 
single--dish data, as if they were measured with the interferometer. The 
combined visibility set is then processed in the standard interferometer 
reduction procedure. In this method the relative weight between single--dish and 
interferometer visibilities has a strong effect on the resulting brightness 
distribution. Furthermore the dirty image calculated from the combined visibility 
set still needs to be deconvolved using CLEAN or other deconvolution algorithms. 
In particular the CLEAN deconvolution algorithm is problematic for short--spacing 
corrected images because it fails on extended structures. We therefore used a 
different approach to combine the PdBI and 30\,m telescope data.\\
The basic idea behind our method is that the missing flux problem only arises 
from an incorrect interpolation of the visibilities in the central part of the 
{\it uv}--plane ($\sqrt{u^2+v^2}<\lambda/S^{proj}_{min}$). Therefore the missing 
flux problem in a finally reduced interferometer map can be solved by replacing 
the questionable part of the {\it uv}--plane by the values calculated from a 
single--dish map with identical extent, grid and flux units. This procedure 
avoids additional CLEANing on the combined data set and the choice of different 
weightings between interferometer and single--dish data. The requirement for the 
single--dish data is the same as described by Vogel \etal\ (\cite{vogel}). For 
the combination we regridded the single--dish data cube to the same spatial and 
velocity grid as the interferometer data. We then converted the flux units 
from $T_{mb}$ to Jy/pixel using the Rayleigh--Jeans approximation, 
$S/beam=\frac{2k}{\lambda^2}T_{mb}$ and $beam=1.133\,f_s^2$ where $f_s$ is the 
FWHM of the 30\,m telescope beam in units of the pixel size. The flux units of 
the cleaned and primary beam corrected interferometer cube were also converted 
to Jy/pixel with $beam=1.133 f_{i,maj}\cdot f_{i,min}$, where $f_{i,maj}$
and $f_{i,min}$ are the FWHM of the major and minor axis of the clean beam in 
units of the pixel size. Furthermore we generated a model for the 30\,m telescope 
main beam and the interferometer clean beam. The 30\,m beam was assumed to be represented
by a circular Gaussian with FWHM = $f_s$. The interferometric beam was described as 
a Gaussian with major and minor axis $f_{i,maj}$ and $f_{i,min}$. 
The normalization of both Gaussians was such that the amplitude of the 
visibility at the origin of the {\it uv}--plane was 1. We then transformed both 
data cubes and the model beams to the {\it uv}--domain using an FFT algorithm. 
The real and imaginary parts of the single--dish data were divided by the amplitudes 
of the model 30\,m beam to deconvolve the single--dish visibilities from the 30\,m 
telescope beam. In order to match the interferometer data the result was then 
multiplied by the amplitudes of the clean beam. At this stage of the combination the real and 
imaginary parts in the single--dish and interferometer data are comparable and the 
central interferometer pixels can be replaced by the single--dish values. The 
part of the {\it uv}--domain to be replaced by the single--dish values in 
general depends on the spacing of the shortest baseline and on the effective 
diameter of the single--dish telescope. For data sets with no overlap in the 
{\it uv}--domain we replaced the part that corresponds to the effective diameter 
of the 30\,m telescope (amplitude of the visibilities for the 30\,m beam 
model $>$ 0.5). Otherwise we selected the part smaller than the shortest 
interferometer baseline. Finally the combined real and imaginary parts were 
transformed back to the image domain and the flux units were converted to Kelvin 
considering that the combined beam is equal to the clean beam.\\
Note that there are no free parameters in the combination of the data sets except for 
the choice which part of the {\it uv}--plane is replaced by the single--dish data. 
The methods require the knowledge of the single--dish beam pattern and the clean 
beam only. A flow chart for the Short--Spacing correction is given in Fig.\ 
\ref{ssc-dia}.

\begin{acknowledgements}
We with to thank the IRAM staff for carrying out the observations and the help provided 
during the data reduction. We thank J. Shen and K.Y. Lo for making available their CO data 
and C. Henkel and A. Heithausen for many fruitful discussions. This research has been 
supported by the Deutsche Forschungsgemeinschaft (DFG) though grant III GK--GRK 118/2.

\end{acknowledgements}

\end{document}